\documentclass[fleqn,usenatbib]{mnras}

\usepackage{newtxtext,newtxmath}

\usepackage[T1]{fontenc}

\DeclareRobustCommand{\VAN}[3]{#2}
\let\VANthebibliography\thebibliography
\def\thebibliography{\DeclareRobustCommand{\VAN}[3]{##3}\VANthebibliography}

\usepackage{graphicx}	
\usepackage{amsmath}	
\usepackage{xcolor}
\usepackage{ulem} 

\newcommand{\app}[1]{Appendix~\ref{sec:#1}}
\newcommand{\eq}[1]{Equation~\ref{eq:#1}}
\newcommand{\fig}[1]{Figure~\ref{fig:#1}}
\renewcommand{\sec}[1]{Section~\ref{sec:#1}}
\newcommand{\tab}[1]{Table~\ref{tab:#1}}

\newcommand{\bluetides}{{\sc BlueTides}}

\newcommand{\eagle}{\mbox{\sc{Eagle}}}

\newcommand{\euclid}{\mbox{\it Euclid}}
\newcommand{\flares}{\mbox{\sc Flares}}

\newcommand{\cloudy}{{\sc cloudy}}

\def\mnras{MNRAS}

\def\apj{ApJ}
\def\apjs{ApJS}
\def\apjl{ApJL}

\def\P3M{P$^3$M}

\title[FLARES IV]{First Light And Reionisation Epoch Simulations (FLARES) IV: The size evolution of galaxies at $z\geq5$}

\author[William J. Roper et al.]{William J. Roper$^{1}$\thanks{E-mail: w.roper@sussex.ac.uk},
Christopher C. Lovell$^{2,1}$,
Aswin P. Vijayan$^{3,4,1}$,
Madeline A. Marshall$^{5, 6}$,
\newauthor 
Dimitrios Irodotou$^{7,1}$, 
Jussi K. Kuusisto$^{1}$,
Peter A. Thomas$^{1}$,
Stephen M. Wilkins$^{1}$
\newauthor 
\\
$^{1}$Astronomy Centre, University of Sussex, Falmer, Brighton BN1 9QH, UK\\
$^{2}$Centre for Astrophysics Research, School of Physics, Astronomy $\&$ Mathematics, University of Hertfordshire, Hatfield AL10 9AB, UK\\
$^{3}$ Cosmic Dawn Center (DAWN) \\
$^{4}$DTU-Space, Technical University of Denmark, Elektrovej 327, DK-2800 Kgs. Lyngby, Denmark \\
$^{5}$National Research Council of Canada, Herzberg Astronomy \& Astrophysics Research Centre, 5071 West Saanich Road, \\ Victoria, BC V9E 2E7, Canada \\
$^{6}$ARC Centre of Excellence for All Sky Astrophysics in 3 Dimensions (ASTRO 3D), Australia \\
$^{7}$Department of Physics, University of Helsinki, Gustaf Hällströmin katu 2, FI-00014, Helsinki, Finland \\
}

\date{Accepted XXX. Received YYY; in original form ZZZ}

\pubyear{2022}

\begin{document}
\label{firstpage}
\pagerange{\pageref{firstpage}--\pageref{lastpage}}
\maketitle

\begin{abstract}
We present the intrinsic and observed sizes of galaxies at $z\geq5$ in the First Light And Reionisation Epoch Simulations (FLARES). 
We employ the large effective volume of FLARES to produce a sizeable sample of high redshift galaxies with intrinsic and observed luminosities and half light radii in a range of rest frame UV and visual photometric bands. This sample contains a significant number of intrinsically ultra-compact galaxies in the far-UV (1500 \AA), leading to a negative intrinsic far-UV size-luminosity relation. However, after the inclusion of the effects of dust these same compact galaxies exhibit observed sizes that are as much as 50 times larger than those measured from the intrinsic emission, and broadly agree with a range of observational samples. This increase in size is driven by the concentration of dust in the core of galaxies, heavily attenuating the intrinsically brightest regions. At fixed luminosity we find a galaxy size redshift evolution with a slope of $m=1.21-1.87$ depending on the luminosity sample in question, and we demonstrate the wavelength dependence of the size-luminosity relation which will soon be probed by the Webb Space Telescope.
\end{abstract}

\begin{keywords}
galaxies: evolution -- galaxies: high-redshift -- galaxies: photometry
\end{keywords}



\section{Introduction}

Galaxy sizes are governed by a range of processes including galaxy mergers, instabilities, gas accretion, gas transport, star formation and feedback \citep{Conselice14}. Studying galaxy sizes helps us understand the interplay between these key astrophysical processes and galactic structure. By extension, understanding how galaxy sizes evolve tells us how these fundamental physical mechanisms, and the interplay between them, change over time. 

At fixed redshift, the size-luminosity relation can be expressed as a power law of the form,
\begin{equation}
    R = R_0 \left(\frac{L}{L_{z=3}^{\star}}\right)^\beta,
\label{eq:size_lumin_fit}
\end{equation}
where $R_0$ is a normalisation factor, $\beta$ is the slope of the size-luminosity relation and $L_{z=3}^{\star}$ is the characteristic ultraviolet (UV) luminosity for $z\sim3$ Lyman-break galaxies (with value $L_{z=3}^{\star}=10^{29.03}$ erg s$^{-1}$ Hz$^{-1}$), which corresponds to $M_{1600} = -21.0$ \citep{Steidel_1999}. As a function of redshift the size evolution can be expressed as
\begin{equation}
    R(z)=R_{0,z=0}(1+z)^{-m}
\label{eq:evo}
\end{equation}
where $R_{0,z=0}$ is another normalisation factor corresponding to the size of a galaxy at $z=0$ and $m$ is the slope of the redshift evolution. In addition to its importance to understanding physical processes, probes of the size-luminosity relation and its evolution are indispensable to our understanding of survey completeness and by extension the luminosity function \citep{Kawamata_2018, Bouwens2021}.

In observations at low redshifts ($z<3$), galaxies have sizes of the order $1-30$ pkpc, with actively star forming galaxies typically larger than their quiescent counterparts \citep{Zhang_2019,Kawinwanichakij_2021}. These galaxies exhibit a positive size-luminosity relation \citep{van_der_Wel_2014,Suess_2019,Kawinwanichakij_2021}, although \cite{van_der_Wel_2014} find a significant number density of compact and massive ($R<2$ pkpc, $M/M_\odot>10^{11}$) galaxies at $z=1.5-3$, whose number density drops drastically by the current day. 

The landscape is different at high redshift where we are primarily probing star forming galaxies. A number of studies using deep Hubble Space Telescope (HST) fields have measured the sizes of $z=6-12$ Lyman-break galaxies \citep{Oesch_2010, Grazian_2012, Mosleh_2012, Ono_2013, Huang_2013, Holwerda_2015, Kawamata_2015, Shibuya2015, Kawamata_2018, Holwerda2020}. In contrast to the low redshift size regime, these studies found bright star forming galaxies with compact half light radii of 0.5-1.0 pkpc. 

There is a growing consensus that the high redshift size-luminosity relation is positively sloped ($\beta>0$), as it is at low redshift, with a range of reported slopes and differing reports of $\beta$'s redshift evolution:
\begin{itemize}
    \item \cite{Grazian_2012} find $\beta=0.3-0.5$ at $z\sim7$.
    \item \cite{Huang_2013} find $\beta=[0.22, 0.25]$ for $z=4$ and $z=5$ respectively.
    \item \cite{Holwerda_2015} find $\beta=0.24\pm0.06$ at $z\sim7$ and $\beta=0.12\pm0.09$ at $z\sim9-10$.
    \item \cite{Shibuya2015} find a redshift independent slope of $\beta=0.27\pm0.01$ in the range $z=0-8$.
    \item \cite{Kawamata_2018} find steeply sloped relations with $\beta=[0.46, 0.46, 0.38, 0.56]$ at $z=[6, 7, 8, 9]$ respectively.
\end{itemize}
Recent lensing studies agree with the steeper slope of \cite{Kawamata_2018}, itself using a sample including lensed sources. \cite{Bouwens2021} find $\beta=0.40\pm0.04$ for a galaxy sample in the redshift range $z\sim6-8$, while \cite{Yang2022} find $\beta=0.48\pm0.08$ for $z\sim6-7$ and $\beta=0.68\pm0.14$ for $z\sim8.5$ (assuming the Bradac lens model \cite{Bradac04}). This steeper slope is driven by compact dim galaxies which are better sampled in lensing studies.

A similar range of results exists within measurements of the redshift dependence of galaxy size at fixed luminosity with slopes in the range $1<m<1.5$ \citep{Bouwens_2004, Oesch_2010, Ono_2013, Kawamata_2015, Shibuya2015, Laporte_2016, Kawamata_2018}. This is consistent with two theoretical scenarios: $m=1$, the expected scaling for systems of fixed mass \citep[e.g.][]{Bouwens_2004}, and $m=1.5$, the expected evolution for systems with fixed circular velocity \citep[e.g.][]{Ferguson_2004, Hathi_2008}.
However, galaxy sizes are not wholly dependent on these theoretical scalings with significant contributions from baryonic processes such as stellar and AGN feedback \citep{Wyithe_2011}.

Simulations provide detailed information on the properties of the underlying components that make up galaxies. From this information we can probe large samples of galaxies with knowledge of the intrinsic physical processes governing their evolution, albeit processes which are themselves dictated by subgrid models which are sensitive to their physical model and parameter assumptions. The intrinsic properties of particles and their spatial distribution can be utilised to measure galaxy properties such as their half mass/light radii at the mass resolution of the simulation without the associated uncertainties inherent in measurements of this kind in observations. Using this fidelity, the size-mass and size-luminosity relations have been probed by many simulations. However, much of this analysis still focuses on comparatively low redshifts. \cite{Furlong_2017} analysed the \eagle\ simulation and found a good agreement with observed trends using intrinsic particle measurements to find a positive ($\beta>0$) size-mass relation which flattens at $z=2$, and an increase in size with decreasing redshift over the range $z=0-2$.

At higher redshift ($z=6$), the \texttt{Simba} simulations \citep{Simba2019} find a positive far UV attenuated size—luminosity relation while showing the dust attenuated size is significantly larger than the intrinsic size, with the magnitude of this increase a function of stellar mass \citep{Wu2020}.  This implies a flatter intrinsic size-luminosity relation at high redshift. This flattened intrinsic size-luminosity relation is particularly evident in the \bluetides\ simulation \citep{Feng2016,Marshall21} which has been used to probe the UV and visual size-luminosity relations with synthetic observations at $z\geq7$. In doing so they find a negative intrinsic size-luminosity relation ($\beta<0$) in the far UV which flips to positive after the inclusion of dust attenuation ($\beta>0$). They also probe the redshift evolution of size, finding a shallow redshift evolution of $m=0.662\pm0.008$ in agreement with the redshift evolution of \cite{Holwerda_2015}. In addition to the higher redshift results derived from \bluetides, the \textsc{Illutris-TNG} simulations have also exhibited a negative size-luminosity relation at $z=5$ \citep{Popping2021}.

The FIRE-2 simulations \citep{Ma_18_size} present a sample of compact galaxies with sizes of 0.05–1 pkpc, in the range $-22<M_{UV}<-7$ at $z=[6,8,10]$. The sizes in this sample are measured from synthetic galaxy images of the intrinsic stellar emission using a non-parametric pixel method, which converts the pixel area containing half the total luminosity to a half light radius. Unlike \cite{Marshall21} this sample exhibits a size-mass relation and B band size-luminosity relation with $\beta>0$. The FIRE-2 galaxy sample extends to galaxies far fainter than those present in other simulated samples, which could explain the differences in size-mass and size-luminosity relations. They also present redshift evolution slopes derived in fixed stellar mass regimes which produce values of $1<m<2$, encompassing many of the observational measurements but extending to more extreme values for the brightest and most massive galaxies.

Clearly there is much work to be done in understanding galaxy size at this epoch, especially with the impending first light of Webb and other next--generation observatories. In this paper we analyse the large sample of galaxies produced by the \flares\ simulations \citep{Lovell2021, Vijayan2020}. \flares\ is uniquely placed to complement previous studies of high redshift galaxy size due to its enormous effective volume, coverage a wide array of environments during the Epoch of Reionisation, and sufficient mass resolution, producing a large and robust galaxy sample. In previous work we have shown that \flares\ reproduces the distributions of stellar mass, star formation rate and UV luminosity up to z ~ 10.

The rest of this article is structured as follows: in \sec{flares} we detail the simulations themselves, in \sec{photo} we detail the methods used to make synthetic photometry and observations, in \sec{sample_methods} we detail the galaxy sample and size measurement methods, and in \sec{size-lumin} we present the results of this analysis of the size-luminosity relation. We present our conclusions in \sec{conclusion}. Throughout this work we assume a Planck year 1 cosmology ($\Omega_{0} = 0.307$, $\Omega_\Lambda = 0.693$, $h = 0.6777$, \cite{planck_collaboration_2014}) and a Chabrier stellar initial mass
function (IMF) \citep{chabrier_galactic_2003}.

\section{First Light And Reionisation Epoch Simulations (\flares)}
\label{sec:flares}

\flares\ is a simulation programme targeting the Epoch of Reionisation (EoR).
It consists of 40 zoom simulations, targeting regions with a range of overdensities drawn from an enormous $(3.2\; \mathrm{cGpc})^{3}$ dark matter only simulation  \citep{barnes_redshift_2017}, which we will refer to as the `parent'.
The regions are selected at $z = 4.67$, which ensures that extreme overdensities are only mildly non-linear, and thus the rank ordering of overdensities at higher redshifts is approximately preserved.  
Regions are defined as spheres with radius 14 cMpc/h, and their overdensities are selected to span a wide range ($\delta=-0.479\to0.970$; see Table A1 of \citealt{Lovell2021}) in order to sample the most under- and over-dense environments at this cosmic time, the latter containing a large sample of the most massive galaxies, thought to be biased to such regions  \citep{chiang_ancient_2013,lovell_characterising_2018}.
These regions are then re-simulated with full hydrodynamics using the \eagle\ model \citep{schaye_eagle_2015,crain_eagle_2015}.

The \eagle\ project consists of a series of hydrodynamic cosmological simulations, with varying resolutions and box sizes.
The code is based on a heavily modified version of \textsc{P-Gadget-3}, a smooth particle hydrodynamics (SPH) code last described in \cite{springel_simulations_2005}.
The hydrodynamic solver is collectively known as \textsc{Anarchy} \citep[described in][]{schaye_eagle_2015,Schaller2015}, and adopts the pressure-entropy formulation described by \cite{Hopkins2013}, an artificial viscosity switch \citep{cullen_inviscid_2010}, and an artificial conduction switch \citep[e.g.][]{price_modelling_2008}. 
The model includes prescriptions for radiative cooling and photo-heating \citep{Wiersma2009a}, star formation \citep{Schaye2008}, stellar evolution and mass loss \citep{Wiersma2009b}, feedback from star formation \citep{DallaVecchia2012}, black hole growth and AGN feedback \citep{springel_blackhole_2005,B_and_S2009,Rosas-Guevara2015}. 
The $z=0$ galaxy mass function, the mass-size relation for discs, and the gas mass-halo mass relation were used to calibrate the free parameters of the subgrid model. 
The model is in good agreement with a number of observables at low-redshift not considered in the calibration \citep[e.g.][]{furlong_evolution_2015,Trayford2015,Lagos2015}. 

\flares\ uses the AGNdT9 configuration of the model, which produces similar mass functions to the fiducial Reference model, but better reproduces the hot gas properties of groups and clusters \citep{barnes_cluster-eagle_2017}. 
It uses a higher value for C$_{\text{visc}}$, a parameter for the effective viscosity of the subgrid accretion, and a higher gas temperature increase from AGN feedback, $\Delta$T. 
These modifications give less frequent, more energetic AGN outbursts. 

The \flares\ simulations have an identical resolution to the 100 cMpc Eagle Reference simulation box, with a dark matter and an initial gas particle mass of $m_{\mathrm{dm}} = 9.7 \times 10^6\, \mathrm{M}_{\odot}$ and $m_{\mathrm{g}} = 1.8 \times 10^6\, \mathrm{M}_{\odot}$ respectively, and has a gravitational softening length of $2.66\, \mathrm{ckpc}$ at $z\geq2.8$. 

In order to obtain a representative sample of the Universe, by combining these regions using appropriate weightings corresponding to their relative overdensity, we are able to create composite distribution functions that represent much larger volumes than those explicitly simulated.
For a more detailed description of the simulation and weighting method we refer the reader to \cite{Lovell2021}.


\subsection{Galaxy Extraction}
\label{sec:extract}

We follow the same structure extraction method as the EAGLE project: this is explained in detail in \cite{Mcalpine_data}. In brief, dark matter overdensities are identified using a Friends-Of-Friends (FOF) approach \citep{davis_evolution_1985} with the usual linking length of $\ell=0.2\bar{x}$, where $\bar{x}$ is the mean inter-particle separation. All other particle types are then assigned to the halo containing their nearest dark matter neighbour. These FOF-halos are then refined to produced self-bound "subgroups" (galaxies) containing both dark matter and baryonic particles using the \textsc{Subfind} algorithm \citep{springel_populating_2001, Dolag2009}. 

The \textsc{Subfind} method involves finding saddle points in the density field in a FOF-halo to identify self-bound substructures. This can lead to spurious oversplitting of extremely dense galaxies where saddle points are misidentified near density peaks. These objects often contain mainly a single particle type and have anomalous integrated properties. Although they make up $<0.1\%$ of all galaxies $>10^8$ M$_\odot$ at $z=5$, we identify and recombine them into their parent structure in post processing. To do this we label a `galaxy' as spurious if it has any zero mass contributions in the stellar, gas or dark matter components. We remove the spurious galaxies from the \textsc{Subfind} catalogue and add their particle properties to the parent `central' subhalo, including the reassigned particles in any integrated quantities. 

In a minority of pathological cases tidal stripping can cause galaxies to exhibit diffuse populations of particles at large radii. Although identified by \textsc{Subfind} as belonging to a galaxy, these distributions can have a large effect on integrated quantities such as the total luminosity and the half light radius. For this reason we adopt a 30 pkpc aperture inline with all \eagle\ and \flares\ papers and calculate all integrated properties within this aperture. This aperture ensures the majority of galaxies have mass distributions which are wholly within this aperture and any erroneous distributions at large radii are omitted.

\section{Modelling Photometry}
\label{sec:photo}

We use the approach presented in \cite{Vijayan2020} (henceforth \flares\-II) to produce resolved galaxy images, both including and excluding the effects of dust. We first produce spectral energy distributions (SEDs) and then apply top hat rest frame UV and visual band filters to extract photometry. As in \flares\-II we focus on the stellar emission, deferring the treatment of accretion onto the super-massive black holes to a future work. However, as will be shown in the coming sections this simplification does not pose a significant challenge to the results of this work. This approach broadly follows \cite{Wilkins2016a,Wilkins2017,Wilkins2018,Wilkins2020}, with modifications to the dust treatment. For a full description of this method and discussion of the free parameters see \flares\-II. What follows is a brief summary of the approach to compute galaxy images. 

\subsection{Spectral Energy Distribution Modelling}

\begin{figure}
	\includegraphics[width=\linewidth]{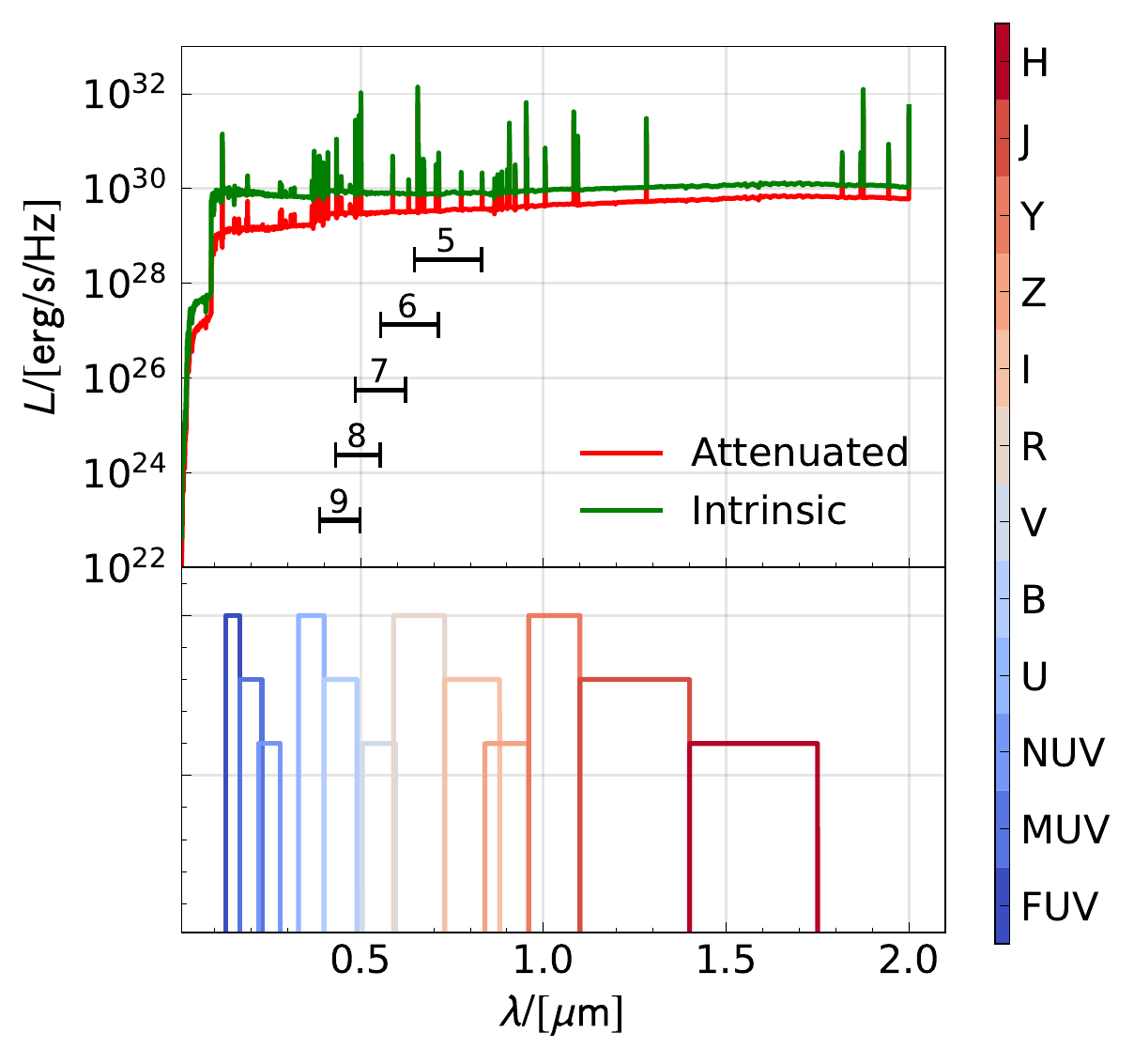}
    \caption{The median rest frame SEDs for all galaxies in all \flares\ regions at $z=5$ with $10^{10} \leq M_\star/M_\odot \leq 10^{11.3}$ produced by \texttt{SynthObs}. The top panel shows the intrinsic stellar SED in green and the dust attenuated SED (including Line of Sight effects) in red. The lower panel shows the rest frame top hat photometric filters used throughout this analysis, plotted with an arbitrary $y$ axis to aid interpretation. The black lines correspond to the location and bandwidth of Webb Space Telescope's near-infrared camera's (NIRCam) reddest wide-band filter (F444W) at the indicated redshifts. This indicates the reddest rest frame bands accessible by Webb at high enough resolution to measure robust sizes with NIRCam (0.062 arcseconds) at $z>5$.}
    \label{fig:example_sed}
\end{figure}

In this work we use the \textsc{SynthObs} module\footnote{\href{https://github.com/stephenmwilkins/SynthObs}{github.com/stephenmwilkins/SynthObs}} to produce synthetic rest frame photometry primarily focusing on a top hat far-UV (1500 \AA) filter with a wavelength range of 1300 \AA $\leq \lambda \leq$ 1700 \AA. We do however calculate results for a range of different filters all shown in the example SED in \fig{example_sed}. Each component of the stellar luminosity can be included independently enabling the probing of both the intrinsic luminosity and the effects of dust extinction. In this section we briefly detail each component.

\subsubsection{Stellar Emission}

For the pure stellar emission we start with a simple stellar population model (SSP) by associating each stellar particle with a stellar SED based on the particle's age and metallicity. As with \flares\-II we use v2.2.1 of the Binary Population and Spectral Synthesis (BPASS) stellar population synthesis (SPS) models \citep{BPASS2.2.1} and assume a \cite{chabrier_galactic_2003} Initial Mass Function (IMF). As shown in \cite{Wilkins2016a,Wilkins2017,Wilkins2018} the resulting luminosities are sensitive to the choice of SPS and IMF used in their derivation. 


\subsubsection{Nebular Emission}

To account for the Lyman continuum emission (LyC) of young stellar populations we associate young stellar particles ($t < 10$ Myr, following the assumption from \cite{Charlot_and_Fall2003} that birth clouds dissipate on these timescales) to a $\mathrm{HII}$ region (or birth cloud). To include the LyC emission for each stellar particle we follow the approach detailed in \cite{Wilkins2020}, in which the pure stellar spectrum is processed with the \cloudy\ photoionisation code \citep{Ferland2017} assuming:

\begin{itemize}
    \item The $\mathrm{HII}$ region's metallicity is identical to the stellar particle's.
    \item Dust depletion and relative abundances from \cite{Gutkin2016}.
    \item A reference ionisation parameter (defined at $t=1$ Myr and $Z=0.02$) of $\log_{10}(U_{S,\mathrm{ref}})=-2$.
    \item A hydrogen density of $\log_{10}(n_{\mathrm{H}}/\mathrm{cm}^{-3})=2.5$.
    \item \texttt{CLOUDY}'s default Orion-type graphite and silicate grains.
\end{itemize}

\subsubsection{Dust Attenuation}
\label{sec:dustatt}

To include the effects of dust attenuation from the ISM we adopt a line of sight (LOS) attenuation model. In this model we treat stellar particles as emitters along a line of sight (in this article we select the z-axis of the simulation) and account for the attenuation due to gas particles which intersect this LOS. Using a LOS approach means stellar emission undergoes spatially resolved attenuation rather than the uniform attenuation of a simple screen model, enabling considerably more robust photometry. 

To do this we find all gas particle SPH kernels which intersect the stellar particle's line of sight and integrate along it to get the metal column density, $\Sigma(x,y)$. We then link this metal column density to the ISM dust optical depth in the V-band (550nm), $\tau_{\mathrm{ISM}}(x,y)$, with a similar approach as in \cite{Wilkins2017}. This gives the expression
\begin{equation}
    \tau_{\mathrm{ISM,V}}(x,y) = \mathrm{DTM} \, \kappa_{\mathrm{ISM}} \, \Sigma(x,y),
\end{equation}
where DTM is the galaxy specific dust-to-metal ratio from the fitting function presented in \cite{Vijayan2019}. This is a function of the mass-weighted stellar age ($t$) and the gas-phase metallicity of a galaxy ($Z$),
\begin{equation}
    \mathrm{DTM}=\mathcal{D}_{0} +(\mathcal{D}_{1} - \mathcal{D}_{0})\left[1-\exp\left(-\alpha Z^{\beta}(t/\tau)^{\gamma}\right)\right],
\end{equation}
where $\mathcal{D}_{0}$ and $\mathcal{D}_{1}$ represent the initial type II SNe dust injection and saturation respectively, and $\tau$ is an estimate of the initial dust growth time-scale after dust injection from type II supernovae but prior to the initiation of dust growth on grains
\footnote{For the parameters of this function we use the best fit values from \cite{Vijayan2019} (see Section 4.1.3 therein for further details): $\mathcal{D}_{0}=0.008$, $\mathcal{D}_{1}=0.329$, $\alpha=0.017$, $\beta=-1.337$, $\gamma=2.122$, and $\tau=5\times10^{-5}[\mathrm{Gyr}]/(\mathcal{D}_{0}Z)$.}
The normalisation factor $\kappa_{\mathrm{ISM}}$ was chosen to match the rest frame UVLF from \cite{Bouwens_2015a} and acts as a proxy for dust properties such as average grain size, shape and composition ($\kappa_{\mathrm{ISM}}=0.0795$). The \flares\ simulations do not inherently model dust production and destruction, thus we have to resort to these data driven proxies.  

In addition to attenuation due to the ISM, young stellar populations ($t < 10$ Myr) are still embedded in their birth clouds and thus need to take into account attenuation due to this cloud. For these young stellar particles we include the additional attenuation expression:
\begin{equation}
   \tau_{\mathrm{BC,V}}(x, y)=\kappa_{\mathrm{BC}}(Z/0.01) ,
\end{equation}
where $Z$ is the metallicity of the young stellar particle and $\kappa_{\mathrm{BC}}$ is another normalisation factor encapsulating the dust properties of the birth cloud, for this we assume a constant value of $\kappa_{\mathrm{BC}}=1$. For stellar particles older than $10$ Myr, $\tau_{\mathrm{BC,V}}(x, y)=0$ and there is no contribution.

We then combine these optical depths in the V-band,
\begin{equation}
    \tau_\lambda= (\tau_{\mathrm{BC,V}}+\tau_{\mathrm{ISM,V}}) \left(\frac{\lambda}{550 \mathrm{ nm}}\right)^{-1},
\end{equation}
yielding an expression for the optical depth at other wavelengths which can be applied to the stellar particle SEDs to account for dust attenuation.


\subsection{Image creation}
\label{sec:image}

We then apply top hat photometric band filters to the SEDs producing photometry for each stellar particle. Using this photometry we produce synthetic observations with a field of view (FOV) of 60 pkpc x 60 pkpc encompassing the entire 30 pkpc aperture in which a galaxy's integrated quantities are measured (corresponding to 9.34, 12.20, and 14.13 arcseconds at $z=5$, $z=8$ and $z=10$ respectively), see \sec{extract}. We adopt a resolution equal to the redshift dependent softening length of the simulation ($s=2.66 / (1 + z)$ pkpc). 

Synthetic images are often created by treating each stellar particle as a 2-dimensional Gaussian kernel. The standard deviation of this kernel can either be defined by the softening length ($\sigma=s$, producing minimal smoothing), the stellar particle's smoothing length ($\sigma=h_{\mathrm{sml}}$, accounting for the local density), or, most often, the proximity to the $N$th neighbouring stellar particle ($\sigma=r_{\mathrm{N}}$) \cite[e.g.][]{Torrey2015,Ma_18_size,Marshall21}. The full image is then a sum over these contributions. In this method an image ($I$) can therefore be expressed mathematically as
\begin{equation}
I_i =  \exp\left(-\frac{(X-x_i)^2 + (Y-y_i)^2}{2\sigma_{i}^2}\right),
\end{equation}
\begin{equation}
I = \sum_{i=0}^{N_\star} \frac{I_i L_i}{\sum_{\mathrm{pix}}I_i},
\end{equation}
where $I_i$ is the smoothed image (kernel) produced for the $i$th stellar particle, $\sigma_i$ is the standard deviation of the $i$th stellar particle's kernel, $X$ and $Y$ are a grid of pixel positions, $x_i$ and $y_i$ are the $i$th stellar particle's x axis and y axis positions in the desired projection, $L_i$ is the luminosity of the $i$th particle, and the sum in the denominator is a sum over all pixels for the $i$th stellar particle to normalise the kernel. 

However, this approach not only differs from the SPH treatment of a stellar particle but is also extremely computationally expensive. Unless artificially truncated a Gaussian kernel encompasses the whole image, leading to insignificant but time consuming calculations. 
In fact, in SPH simulations a stellar particle is treated as a representation of a fluid with the full extent of the stellar population described by a spline kernel with a definitive cut off where the kernel falls to 0 \citep{Borrow2021}. Using a spline kernel based approach is not only a better representation of the underlying simulation's treatment of stellar particles but also greatly reduces the size of the computation by limiting the number of pixels computed per stellar particle.

For these reasons we implement a method of smoothing employing the SPH kernel used in the simulation to describe a stellar particle's `extent'. In the \textsc{Anarchy} SPH scheme, used in the EAGLE model \citep{schaye_eagle_2015}, this kernel is the $C_{2}$ Wendland kernel \citep{wendland_piecewise_1995, Dehnen2012}. We therefore adopt this kernel in this work, but note that for other simulations the kernel corresponding to that particular simulation should be used to maximise the fidelity of this method.

As with the Gaussian approach, an image can be described as a sum over kernels; unlike the Gaussian approach however, the spline kernels are necessarily 3-dimensional and need projecting into the $x-y$ plane. To achieve this we calculate the spline kernels on a voxel grid and sum over the $z$-axis,  
\begin{equation}
I =  \sum_{z\mathrm{-axis}}\sum_{i=0}^{N_{\star}} \frac{K_i}{\sum_{\mathrm{vox}} K_i}L_i,
\end{equation}
where each stellar particle's kernel ($K_i$) is now
\begin{equation}
K_i =  \frac{21}{2\pi} \frac{w_i}{h_{\mathrm{sml}}^3},
\end{equation}
with the kernel $w_i$ given by
\begin{equation}
        w_i(q_i=r/h_i) = 
\begin{cases}
    \left(1-q_i\right)^{4}\left(1+4q_i\right), & q_i \leq 1\\
    0,              & q_i > 1
\end{cases},
\end{equation}
where $r$ is the distance between the particle and any given voxel within the kernel. 

To compute this kernel efficiently we employ a KD-Tree algorithm, building a tree based on voxel coordinates. We query the tree for all non-zero pixels where the distance between the pixel and the stellar particle ($r$) is less than the limits of the smoothing kernel (here $r<h$), greatly reducing the computation from $O(N_\star N_{\mathrm{pix}})$ in the Gaussian case to $O(N_\star N_{\mathrm{vox}(r<h)})$ using the more representative spline approach.

\begin{figure*}
	\includegraphics[width=\linewidth]{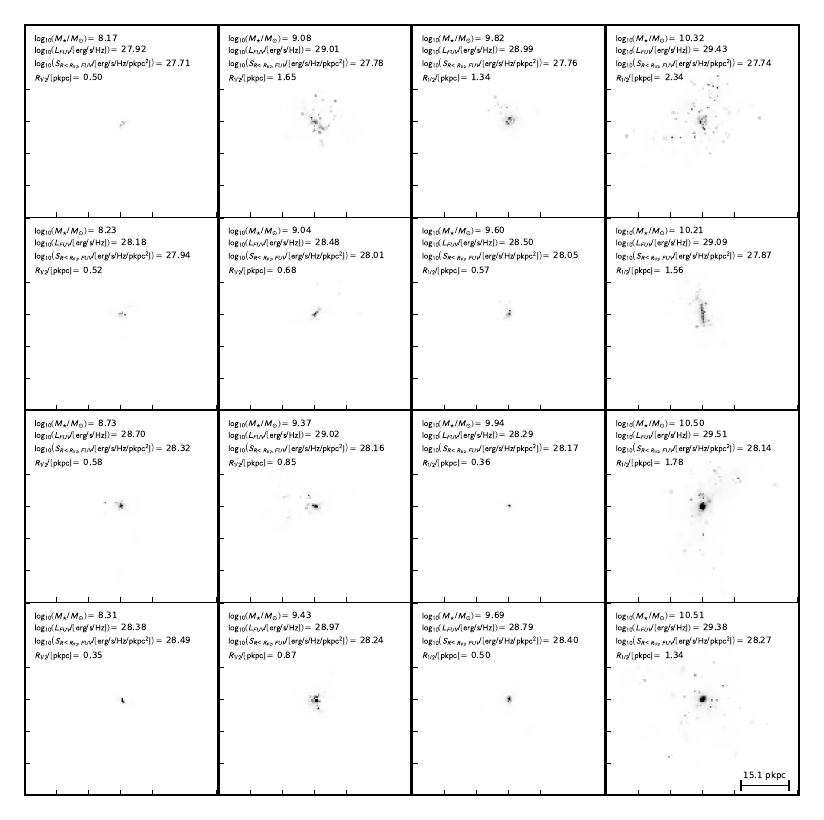}
    \caption{A subset of $z=5$ synthetic far UV galaxy images computed using the method outlined in \sec{image}. Each panel is the full 60 pkpc x 60 pkpc FOV for each galaxy. Galaxies increase in mass left to right and increase in central surface density top to bottom. The pixel values of these images are linearly normalised across all panels with their mass, luminosity, central surface density and half light radius included in each panel. The galaxies included in this subset were randomly selected from each mass and central surface density bin, even so they display the variety of morphologies already present by $z=5$ in \flares.}
    \label{fig:example_img}
\end{figure*}

In \fig{example_img} we present a grid of randomly selected galaxy images in the far-UV filter along with their stellar mass (derived by summing the underlying particle distribution), luminosities, central surface densities and half light radii measured including the effects of dust. It should be noted that throughout this analysis we do not rotate galaxies, instead adopting their existing orientation in the box to emulate the stochastic viewing angles of galaxies in the real Universe. Henceforth, all analysis derived from images will use this method of stellar particle smoothing (implemented from \sec{pixel} onwards), unless explicitly stated otherwise. In \app{smooth} we present comparisons between the Gaussian and spline approach for this simulation. 

\section{Galaxy Selection and Size Measurement}
\label{sec:sample_methods}

In this section we describe our galaxy sample, and describe the two measurement methods used to derive sizes. 

\subsection{Extracting the galaxy sample}
\label{sec:sample}

To ensure all galaxies in the sample have enough particles to be considered morphologically resolved, we omit all subgroups with fewer than 100 stellar particles ($N_\star<100$). We apply a 95 per cent completeness criterion, dividing the sample of galaxies into those above and below the completeness limits in mass and luminosity. These completeness limits are given by the mass and luminosity at which the galaxy sample is missing 5 per cent due to galaxies having $N_\star<100$. We adopt 95 per cent complete rather than 100 percent complete to avoid the luminosity threshold being defined by anomalously bright galaxies with $N_\star<100$. These limits are presented in \tab{complete} at each redshift for the far UV band. This ensures we present results motivated by a complete galaxy sample. We nonetheless present the incomplete sample at low opacity in all scatter plots for context.

\begin{table*}
\begin{center}
\begin{tabular}{ c | c | c | c }
 \hline
 Redshift ($z$) & $\log_{10}(M/M_\odot)$ & $\log_{10}(L_{\mathrm{int}}/[\mathrm{erg} / \mathrm{s} / \mathrm{Hz}])$ & $\log_{10}(L_{\mathrm{att}}/[\mathrm{erg} / \mathrm{s} / \mathrm{Hz}])$ \\ \hline
 12 & 8.16 & 28.60 & 28.43 \\  
 11 & 8.15 & 28.55 & 28.42 \\  
 10 & 8.15 & 28.52 & 28.39 \\  
 9 & 8.14 & 28.46 & 28.34 \\  
 8 & 8.13 & 28.40 & 28.28 \\  
 7 & 8.13 & 28.31 & 28.19 \\  
 6 & 8.12 & 28.24 & 28.12 \\  
 5 & 8.11 & 28.16 & 28.03 \\
 \hline
\end{tabular}
\caption{The mass and luminosity 95 per cent completeness limits for the galaxy sample in each redshift bin. The mass limits are consistent across all bands, but the luminosity limits are band specific. Here we present the far-UV (FUV, 1500 \AA) limits focused on for the majority of the analysis presented in this article.}
\label{tab:complete}
\end{center}
\end{table*}

We further distinguish between 2 morphological populations by applying a threshold derived from the intrinsic size-luminosity relation of $S \geq 10^{29}$ erg s$^{-1}$ Hz$^{-1}$ pkpc$^{-2}$ to their central surface flux density (i.e. the surface flux density within the half light radius). This threshold splits the sample into a population of centrally compact galaxies and a population of diffuse galaxies; in subsequent plots we will denote the compact population by coloured hexbins and the diffuse population by greyscale hexbins.

\begin{figure}
	\includegraphics[width=\columnwidth]{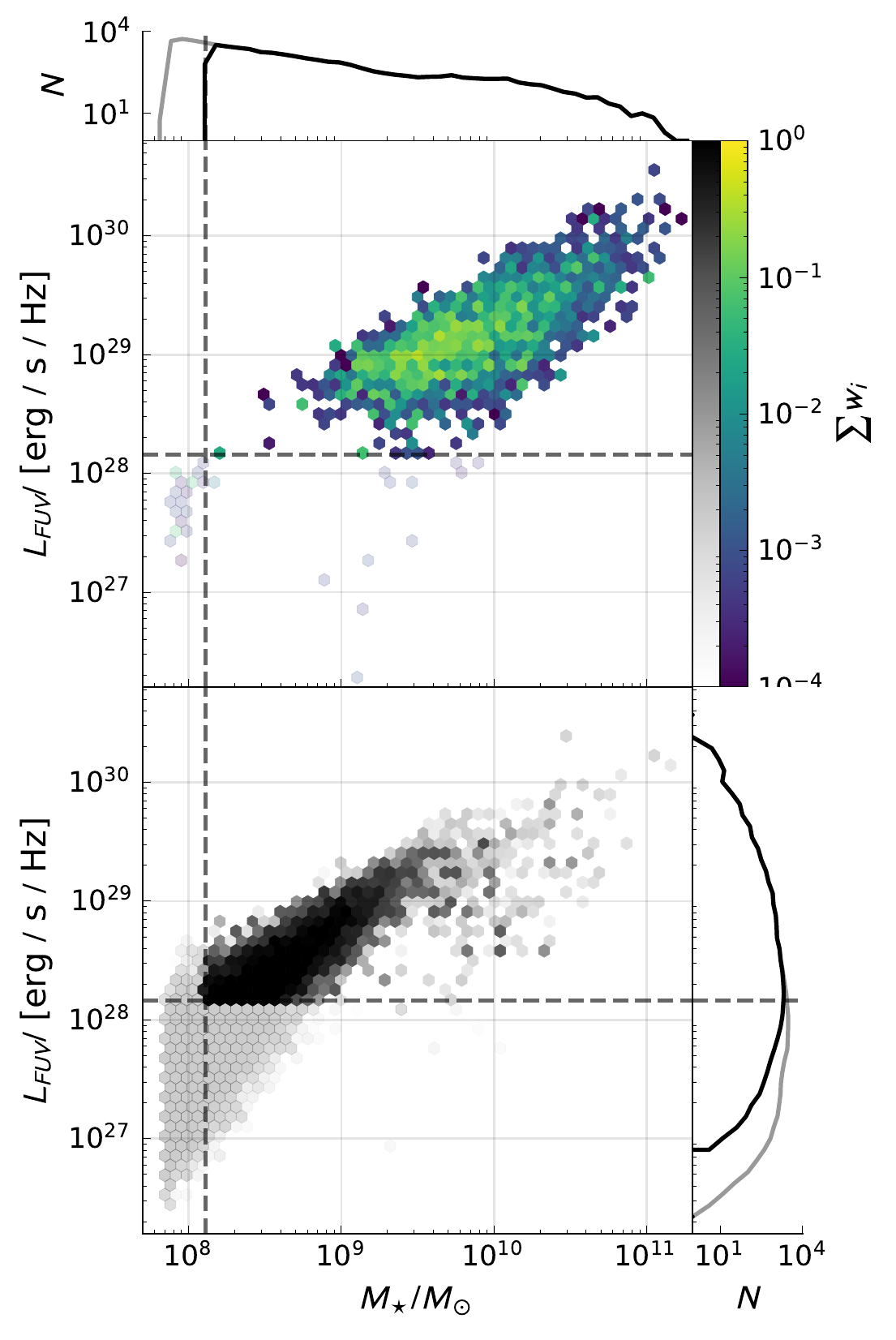}
    \caption{The intrinsic mass-luminosity relation at $z=5.0$. The top panel of coloured hexbins are the galaxies in the compact population and the lower panel of greyscale points represent galaxies in the diffuse population, as described in \sec{sample}. The dashed lines show the completeness limits for the galaxy sample with those galaxies that fall outside this completeness threshold denoted by low opacity. Each hexbin is coloured by the weighted number density of galaxies, using the \flares\ region weighting scheme. The histograms on each axis show the total distribution of galaxies in both the compact and diffuse population along each axis, with the grey line showing all galaxies and the black line showing those with $N_\star\geq100$.}
    \label{fig:int_lumin_mass}
\end{figure}

This division of the galaxy sample is shown in the mass-luminosity relation in \fig{int_lumin_mass} at $z=5$; here we have adopted the previously described colouring and have used opacity to distinguish the complete and incomplete populations. The dashed lines denote the completeness limits in mass and luminosity. The histograms on the axes show the galaxy distribution along each axis with the full galaxy population in grey and galaxies with $N_\star\geq100$ shown in black. 

All following plots will follow these plotting conventions, with greyscale colours denoting the diffuse galaxy distribution and coloured hexbins denoting the compact population (as defined by their central surface density). The hexbins themselves indicate the weighted number density of galaxies, using the weights derived in \cite{Lovell2021}. All fits are performed on the complete sample. This division of the galaxy sample leads to:
\begin{itemize}
    \item 50238 galaxies in the sample with more than 100 stellar particles (25556, 2863 and 492 at $z=5$, $z=8$ and $z=10$ respectively).
    \item 7172 in the compact population with more than 100 stellar particles (2701, 696 and 240 at $z=5$, $z=8$ and $z=10$ respectively).
    \item 43066 in the diffuse population with more than 100 stellar particles (22855, 2167 and 252 at $z=5$, $z=8$ and $z=10$ respectively).
    \item 31697 galaxies in total above the completeness limit (16238, 1700 and 273 at $z=5$, $z=8$ and $z=10$ respectively).
\end{itemize}




\subsection{Size measurement methods}
\label{sec:measure}

There are a myriad of methods used to define the sizes of galaxies present in the literature including S\'ersic profile fitting \citep{Sersic1963, Sersic1968}, curves of growth \citep[e.g.][]{Ferguson_2004, Bouwens_2004, Oesch_2010}, Petrosian radius \citep{Petrosian1976},
and simulation specific methods, that use the particle distribution to find the radius enclosing a percentage of the total mass/luminosity. 

Each measurement method introduces its own dependencies and challenges. In this section we detail and compare the two methods utilised in this analysis: a particle based method, and a non-parametric pixel based method \citep[e.g.][]{Ribeiro2016, Ma_18_size, Marshall21}. We neglect curves of growth, Petrosian radius and S\'ersic profiles entirely; at these redshifts the clumpy nature of galaxies, particularly at lower masses \citep{Jiang_2013, Bowler2016}, make these methods unreliable. Throughout this work we use $R$ to refer to the half light radius (size) of a galaxy.

\subsubsection{Particle based method}

We take the underlying particle distribution within a 30 pkpc aperture and find the radius of the particle bounding half the total luminosity inside this aperture. We then interpolate around this initial measurement to better sample the radial density profile, mitigating it's discretisation into individual, comparatively low resolution, particles. 

It should be noted that this measurement method is sensitive to the chosen galactic centre; in this work we use the centre of potential calculated by SUBFIND. Other choices, such as the centroid, can give different results for diffuse and irregular structures since the centre of potential may be located within one of the clumps, which may not necessarily lie in the centre of the galaxy. This offset centre leads to larger size measurements, as the majority of the stellar material of the galaxy is offset from the centre from which the radius is measured.

In all plots including this measurement we take the luminosity to be the sum of each individual particle's luminosity within the aperture, neglecting any smoothing over the SPH kernel. 




\subsubsection{Pixel based method}
\label{sec:pixel}

In the non-parametric pixel approach, the pixels of the image are ordered from most luminous to least luminous. We then find the pixel area containing half the total luminosity before converting to a radius assuming a circular area, $R=\sqrt{A/\pi}$, and then interpolating around this radius as in the particle method. Unlike the particle method this method of measurement has a minimum possible size where half the total luminosity falls within a single pixel, resulting in a radius of $R_{\mathrm{min}}=\sqrt{A_{\mathrm{pix}}/\pi}$ before interpolation between 0 and $R_{\mathrm{min}}$. The interpolation here allows for the measurement of half light radii smaller than a single pixel, however this does not remove the limitation caused by the finite pixel resolution.

This method is particularly robust at high redshifts, where the independence from a centre definition and non-contiguous size definition better encapsulate the morphology of clumpy structures.

In all plots using this measurement we present the luminosities as detected from the image, i.e. the sum of all pixels within the FOV. This can subtly differ from the particle luminosities where a particle's kernel extends beyond the bounds of the FOV, spreading the particles light outside the image in contrast to the particle based method.

\subsubsection{Comparing particle and pixel methods}

\begin{figure}
	\includegraphics[width=\columnwidth]{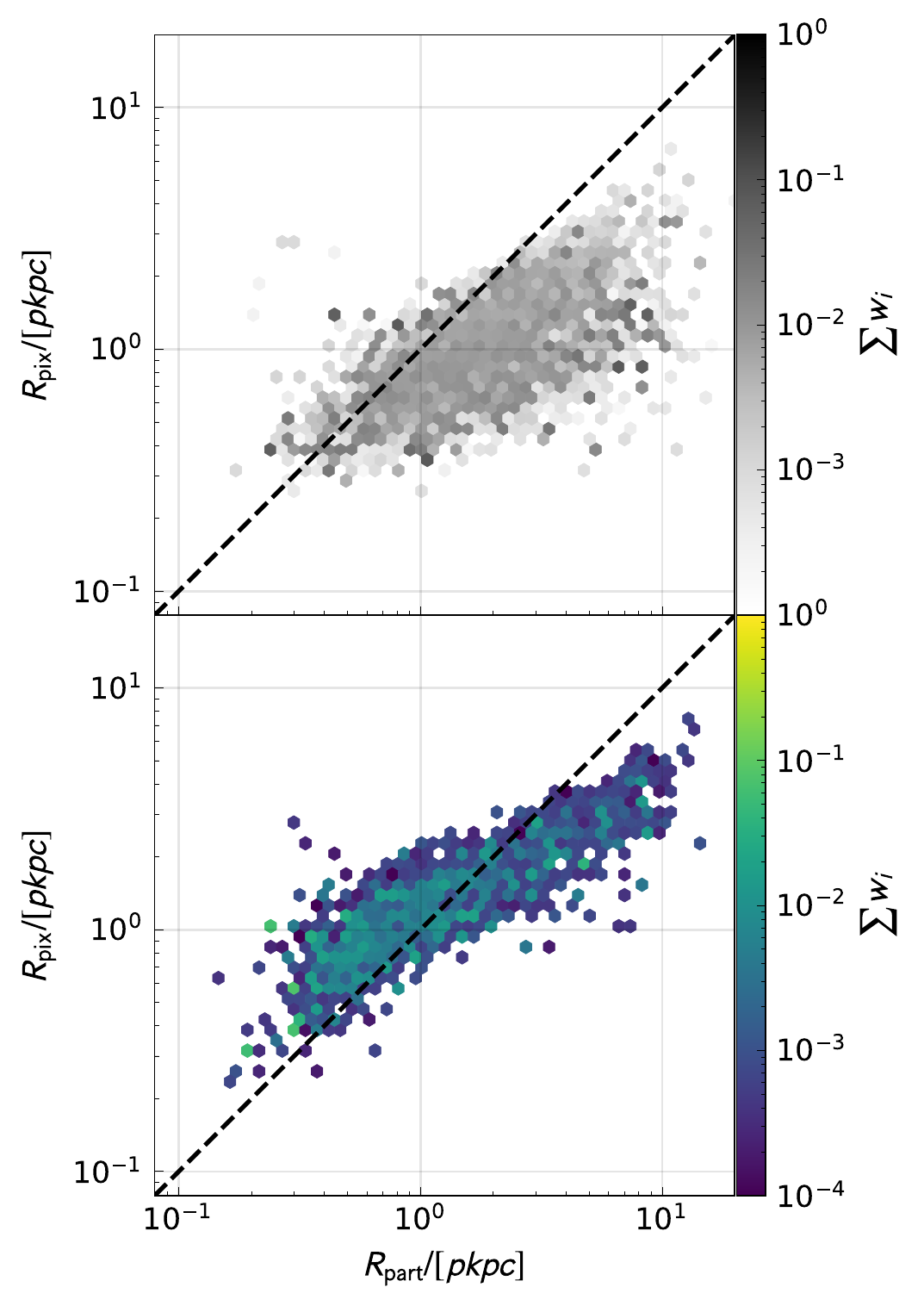}
    \caption{A comparison between the dust attenuated half light radii of galaxies at $z=5$ yielded by the particle measurement method (x-axis) and the pixel method (y-axis). The upper panel of greyscale points show the diffuse galaxy population while the lower panel of coloured points show the compact galaxy population. The dashed black line corresponds to a 1:1 relationship. Each hexbin is coloured by the weighted number density of galaxies, using the \flares\ region weighting scheme.}
    \label{fig:size_method_comp}
\end{figure}

In \fig{size_method_comp} we present a comparison of these methods for the sizes of all galaxies at $z=5$ using their intrinsic luminosities. For the compact galaxies (colour) we see a reasonable correspondence between the two methods with a scatter around the 1:1 relation. However, as the size of a galaxy increases the particle method begins to produce larger sizes than the pixel method due to a combination of centring effects and luminous structures within the outskirts of galaxies, such as those shown in a number of panels in \fig{example_img}. Conversely, for the smallest galaxies, the pixel size is larger than the particle size; this is a manifestation of the stellar particle smoothing used in the creation of the images, where light concentrated in densely packed particles is smoothed over a larger pixel area. 

For the diffuse (greyscale) population the scatter is more pronounced and extends towards larger particle values across the full range of sizes. This is because of the aforementioned strength of the pixel method when it comes to clumpy diffuse structures and the issue of defining a centre for these structures in the particle method. The size floor is also evident in the smallest galaxies in the diffuse (and incomplete) sample where a single pixel contains half the total luminosity of the dim galaxy. 

\section{Size-Luminosity relations}
\label{sec:size-lumin}


Here we present results for the sizes of galaxies in the epoch of reionisation. All plots that compare to observational quantities are derived from the pixel measurement method (\sec{pixel}) measured from the synthetic images detailed in \sec{image}. Intrinsic properties such as the intrinsic size-luminosity relation (\sec{intrinsic}) and half dust radius (\sec{dustdist}) are measured using the particle method to focus on the intrinsic nature of these properties.

\subsection{Intrinsic UV size-luminosity relation}
\label{sec:intrinsic}

\begin{figure}
	\includegraphics[width=\linewidth]{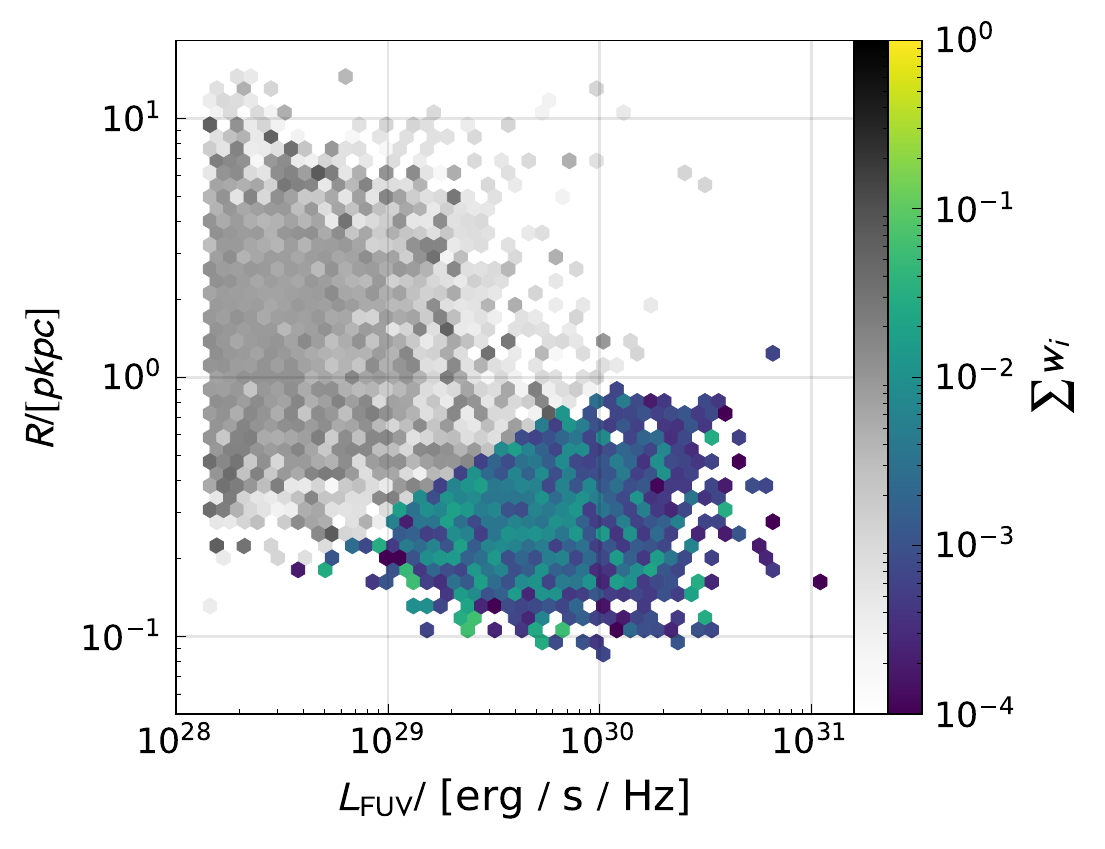}
    \caption{The intrinsic UV size-luminosity relation at $z=5.0$, measured using the particle method. Showing the dimmer diffuse population in greyscale and the bright compact population in colour. The hexbins are coloured by the sum of \flares\ region weightings for each individual galaxy, making each hexbin a weighted number density in the UV size-luminosity plane.}
    \label{fig:int_hlr}
\end{figure}

\begin{figure*}
	\includegraphics[width=\linewidth]{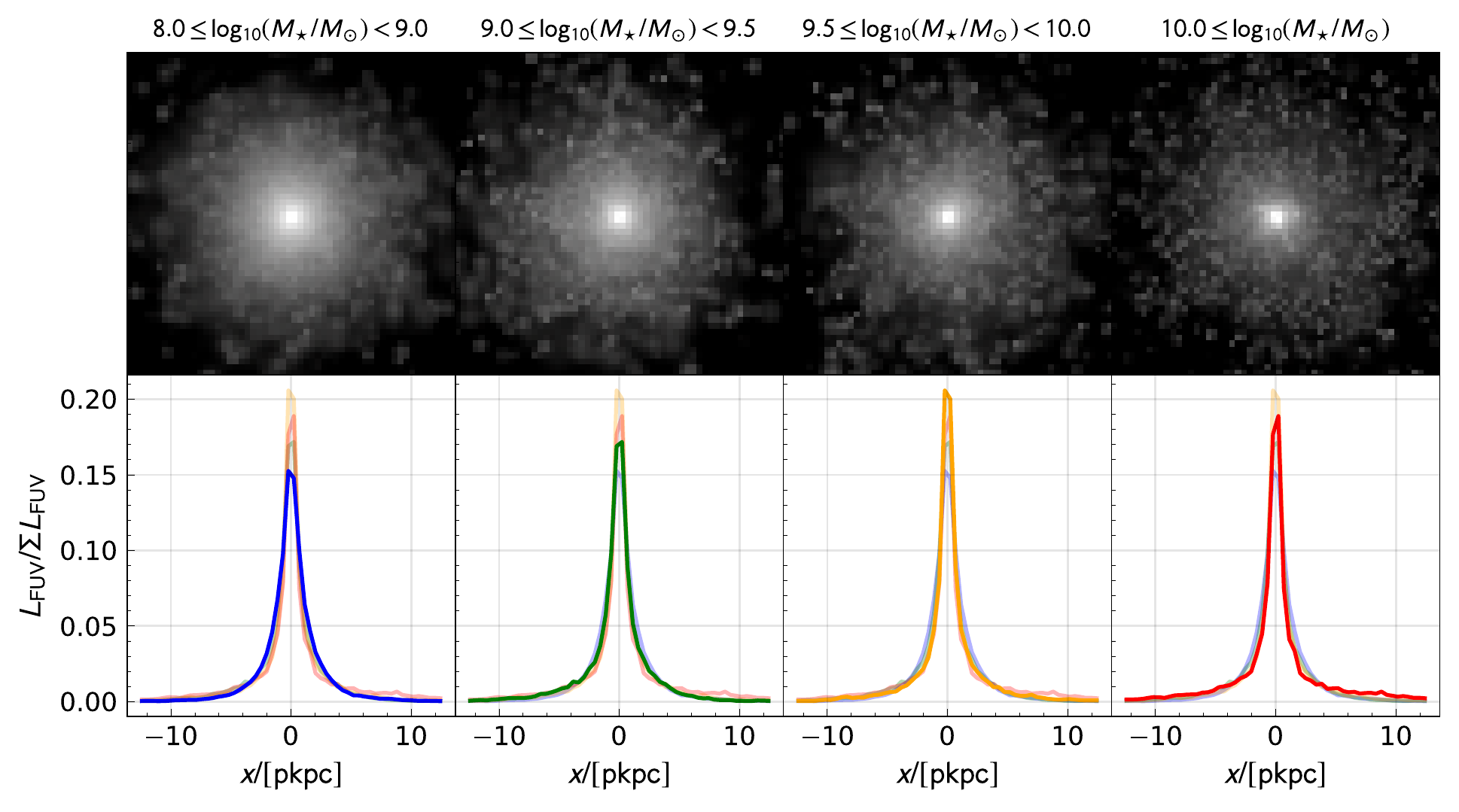}
    \caption{The upper panel contains stacked individually log-scaled images of the intrinsic luminosity of every galaxy in our complete galaxy sample at $z=5$. These stacks cover the central $\sim24$ pkpc of the image and are split into mass bins (increasing left to right). The lower panels show 1-dimensional profiles of these stacks. The luminosity on the y-axis of these profiles is normalised to the sum of each stacked image. In each panel all profiles are plotted with the curves corresponding to the other panels plotted in low opacity to aid interpretation.}
    \label{fig:int_stack}
\end{figure*}

Although not probed in observations, we can use the intrinsic UV size-luminosity relation to trace the underlying stellar population in galaxies. \fig{int_hlr} shows this relation at $z=5$ for the particle measurements. This shows two surprising features: 2 distinct populations, and a clear negative slope to the intrinsic size-luminosity relation. 

Although the negative slope of the intrinsic size-luminosity relation is somewhat counter intuitive, it has been seen at these redshifts in other recent simulations, particularly in \bluetides\ \citep{Marshall21} with a negative size-mass relation at $z=7$ and \textsc{Illutris-TNG} \citep{Popping2021} with a negative observed-frame 850 $\mu$m size-mass relation at $z=5$. Indeed, there are also hints in observations with evidence for a constant dependence between galaxy size and mass \citep{Lang_2014, Mosleh_2020}. 

Here the division in central surface density is particularly evident. In terms of luminosity we have one dim ($L\lesssim10^{29}$ erg s$^{-1}$ Hz$^{-1}$) and more diffuse population, and one bright ($L\gtrsim10^{29}$ erg s$^{-1}$ Hz$^{-1}$) and compact ($R_{1/2}\lesssim 1$ pkpc) population.

We will present our investigation into the physical mechanisms governing this bi-modality in detail in an upcoming paper, but for context in \flares\ (the \eagle\ model):
\begin{itemize}
    \item At $z\gtrsim5$, galaxies that reach $M/\mathrm{ M}_\odot\gtrsim10^9$ develop extremely dense cores and begin a spike in core star formation at high stellar birth densities.
    \item This begins to seed the gas in the galaxy's core with metals, increasing the effectiveness of metal line cooling, inhibiting stellar and AGN feedback, and further driving star formation. 
    \item This overcooling causes a feedback loop of star formation in the galaxy's core, allowing the galaxy to become massive and ultra compact during this early epoch. 
    \item While this process takes place in the galaxy's core the galaxy accretes an extended gas distribution up to 100 times larger than the stellar distribution. Due to the high densities in the core, stellar feedback is unable to mix the core's metals into this surrounding gas distribution. This lack of metals inhibits cooling and leaves the extended gas distribution unable to efficiently form stars.
    \item At $z\lesssim4$ the extended gas distribution reaches the density and metallicity necessary for efficient star formation. This is facilitated partly by their own collapse and partly due to the growing efficiency of stellar and AGN feedback \citep{crain_eagle_2015}, mixing metals from the core into the surroundings. This extended star formation manifests as an increase in galaxy size at late times, yielding the size distribution we see at the present day.
\end{itemize}

In the upper panel of \fig{int_stack} we present a stack of the central intrinsic emission of all galaxies at $z=5$ in \flares\ (irrespective of completeness) split into mass bins of $\log_{10}(M/M_\odot)=[8-9, 9-9.5, 9.5-10, >10]$. This qualitatively shows how the negative gradient in the size-luminosity relation translates to the compactification of a galaxy's intrinsic emission in relation to a galaxy's mass. In the lower panel of \fig{int_stack} we plot 1-dimensional profiles of the stacked mass bin images to explicitly show the compactification. As with the stacked images, the profiles exhibit a narrowing and increasing central concentration with increasing mass. The overcooling begins to take effect between the left most mass bin ($10^{8}<M/M_\odot<10^{9}$) and the next mass bin of $10^{9}<M/M_\odot<10^{9.5}$. At this crossover between regimes there is a narrowing of the profile and stronger concentrated peak, which becomes more peaked as the mass increases. The growth of this central peak then drops off in the final mass bin due to an increased contribution by the wings of the profile; galaxies in this mass bin exist in the most dense environments and thus include more luminous substructure at large radii.

\subsection{The effects of dust}
\label{sec:dust}

We now move on from the intrinsic size-luminosity relation to discuss the effects of dust on the observed UV size-luminosity relation. All plots from this point on will present the pixel measured sizes unless explicitly stated otherwise.

\subsubsection{The distribution of dust}
\label{sec:dustdist}

\begin{figure}
	\includegraphics[width=\columnwidth]{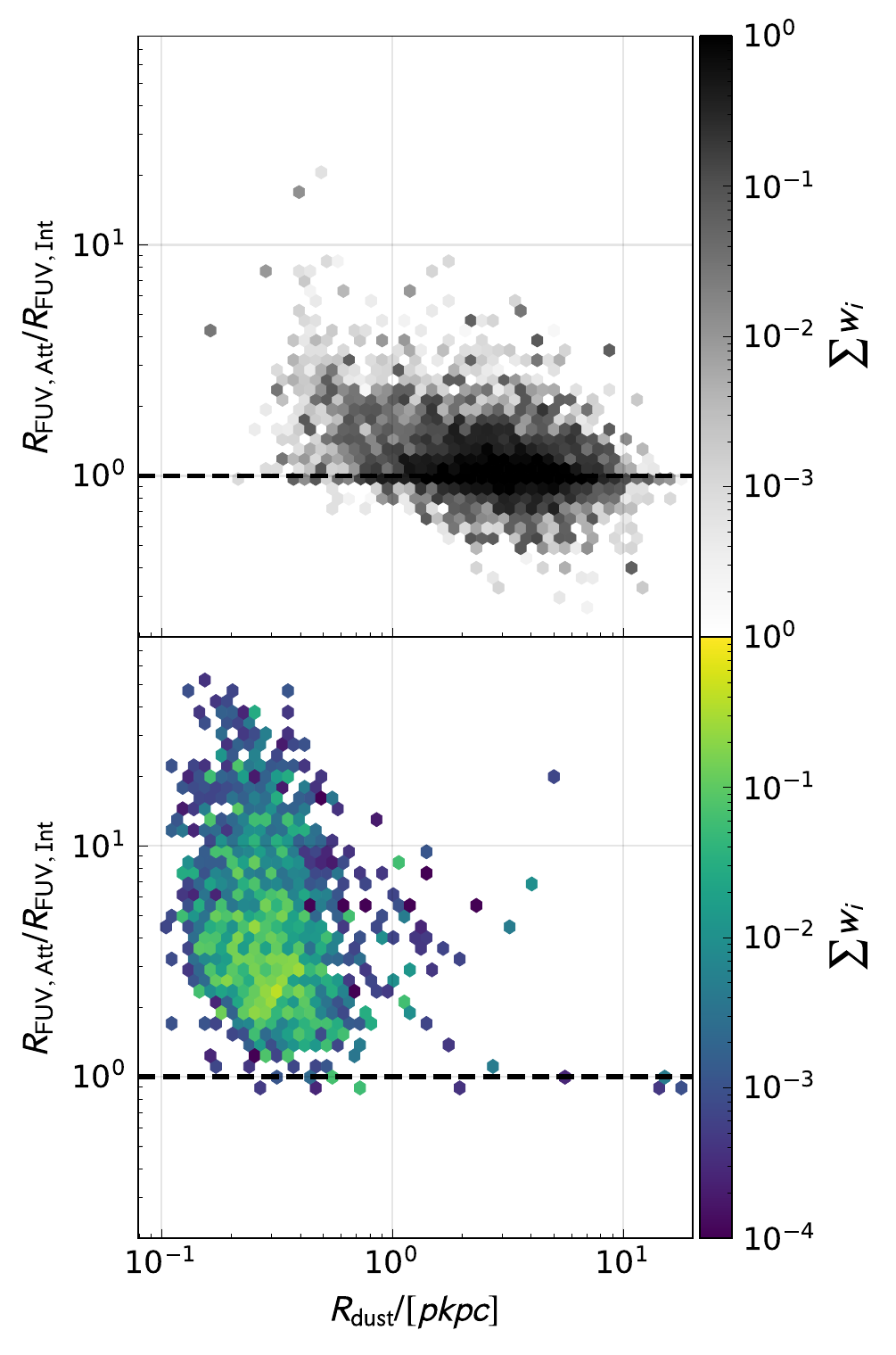}
    \caption{The ratio between dust attenuated and intrinsic size as a function of the half dust radius (the radius enclosing half the mass in gas-phase dust) for all galaxies at $z=5$, computed using the particle method. Once again, the galaxy sample is divided into the diffuse population (upper, greyscale) and the compact population (lower, coloured) and the hexbins are coloured by the cumulative weighting of each galaxy within a hexbin.}
    \label{fig:HDR}
\end{figure}

Dust attenuates the intrinsic stellar emission making observations of the pure stellar emission impossible. The affect this obfuscation will have on the measured size of a galaxy is sensitive to the spatial distribution of dust in a galaxy: a uniform screen would have no discernible effect on the size, whereas any concentration of dust in a particular region will have important consequences for the spatial distribution of observed stellar emission, and therefore perceived size. 

We probe the underlying dust distribution in these galaxies by calculating the half dust radius (i.e. the radius enclosing half the mass in gas-phase dust). To calculate the gas-phase dust mass we use the metallicity of each gas particle and multiply by the galaxy specific $\mathrm{DTM}$ (described in \sec{dustatt}) to get the dust mass of each gas particle. 

\fig{HDR} shows the ratio between attenuated and intrinsic particle based sizes as a function of this half dust radius at $z=5$. Galaxies in the compact population (coloured hexbins) have dust distributions with $R_{1/2, \mathrm{dust}} \lesssim 1$ pkpc and $R_{\mathrm{att}}/R_{\mathrm{int}}\gtrsim1$. This indicates that, in the compact galaxy sample, not only is the distribution of dust highly concentrated in the core of the galaxy, but the more concentrated the dust, the larger the increase in observed size due to the attenuation of the galaxy's bright core\footnote{This strong attenuation of the core justifies the omission of the AGN contribution to the UV luminosity. We have confirmed the AGN contribution is heavily attenuated at these wavelengths, in fact only a handful of galaxies in the sample have AGN that are comparable to their host galaxy in the UV luminosity.}. 
With the central regions strongly attenuated, the more extended regions are able to contribute more to the total luminosity of the galaxy, increasing the perceived size. In the most extreme cases, galaxies can appear $\sim50$ times larger when including dust attenuation.

The vast majority of the diffuse galaxy population (greyscale) also have diffuse dust distributions ($R_{1/2, \mathrm{dust}} > 1$ pkpc) and exhibit a more conservative increases in size between intrinsic and attenuated size. Compared to the compact population, the more diffuse dust distributions (and galaxies) have a flatter relation between the ratio of sizes and half dust radius. Both the smaller increase in size and the flattening of this relation can be explained by a more uniform distribution of dust in these diffuse clumpy structures\footnote{Those galaxies in the diffuse population that do not follow this trend (i.e. exhibit large increases in size with the inclusion of dust and have compact dust distributions) are galaxies very close to the central surface flux density threshold used to split the populations.}.

Galaxies that fall below the dashed line, indicating a ratio of 1, represent a decrease in size with the inclusion of dust effects. These are instances where the dust is more uniformly distributed, and results in greater attenuation of their extremities, driving down the apparent size.

\subsubsection{The Observed UV size-luminosity distribution}

\begin{figure*}
  \includegraphics[width=\linewidth]{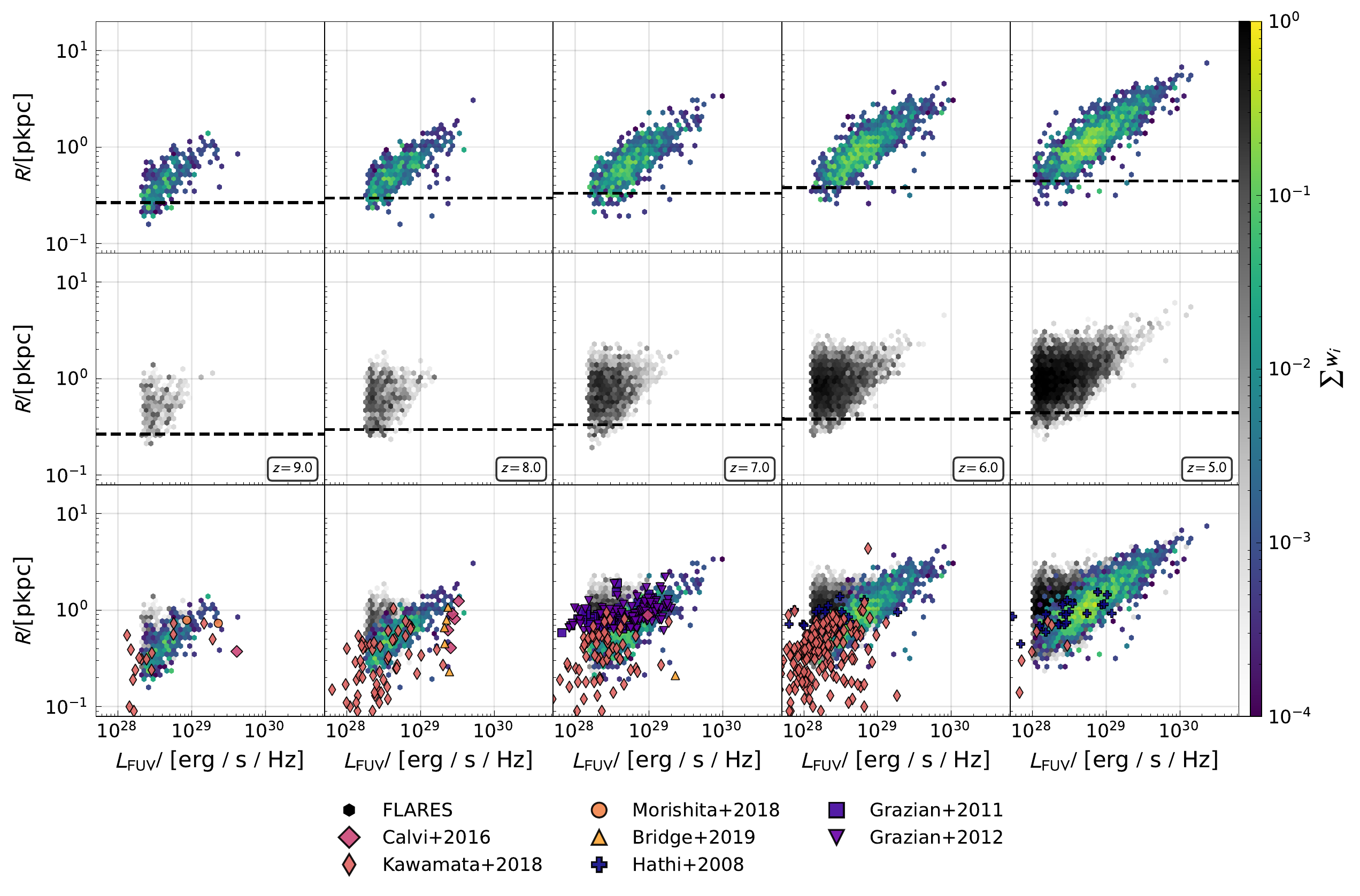}
  \caption{The attenuated far-UV (1500 \AA) size-luminosity relation measured using the pixel method. The hexbins are again coloured by the weighted number density. The galaxy sample is divided into the compact galaxy population (top row, colour) and the diffuse galaxy population (middle row, greyscale). The dashed line shows the pixel resolution of the images used to make the \flares\ measurements. Galaxies can fall below this line due to the interpolation used in the calculation of the pixel half light radius. The bottom row contains both galaxy populations with a comparison to high redshifts observations using the Hubble space telescope \citep{Hathi_2008, Grazian2011, Grazian_2012, Calvi_2016, Kawamata_2018, Morishita_2018, Bridge_2019}.}
  \label{fig:HLR_with_dust_obscomp}
\end{figure*}

The negative gradient in the intrinsic size-luminosity relation presented in \fig{int_hlr} is in direct conflict with observational results which necessarily include the effects of dust attenuation \citep[e.g.][]{Hathi_2008, Grazian2011, Grazian_2012, Shibuya2015, Calvi_2016, Kawamata_2015, Kawamata_2018, Morishita_2018, Bridge_2019, Bouwens2021, Yang2022}. However, in \sec{dustdist} we have shown that the inclusion of dust attenuation can result in large increases in size for the most intrinsically compact galaxies. Ascertaining if this effect is enough to yield sizes in line with observations is imperative to probe the validity of the negative intrinsic size-luminosity relation, and thus the physical models used in \flares. 

To compare to the observed results we use the method detailed in \sec{image} for synthetic image creation and the pixel measurement method (\sec{pixel}) to produce the observed size-luminosity relation and compare to a wide array of observations in integer redshift bins from $z=5-9$. This observed size-luminosity relation is shown in \fig{HLR_with_dust_obscomp}. 

Evidently, the concentration of dust in compact cores and increase in size between intrinsic and attenuated sizes, detailed in \sec{dustdist}, has completely reversed the slope of the size-luminosity relation relative to the intrinsic relation. 

Focusing on the high central surface density distribution (coloured hexbins), beyond the positive relation between size and luminosity, we can already see a power law relation with minimal scatter. This scatter is increased for the diffuse, low central surface density population (greyscale hexbins), particularly for low luminosity galaxies which exhibit a large range of sizes at fixed luminosity. We can also see that the \flares\ galaxy sample extends to larger sizes and higher luminosities than the observed results, this is because of \flares's focus on rare and extreme environments where the most luminous galaxies reside.

There is a fair agreement between the scatter of observational measurements and the \flares\ distribution with the exception of galaxies in the \cite{Kawamata_2018} (lensed) sample which have sizes smaller than the resolution of \flares. Particularly evident when comparing the \flares\ and observational scatter are the \cite{Grazian2011} and \cite{Hathi_2008} (dropout selected) points at $z=7$ and $z=6$, respectively, with similar normalisation to the low central surface density galaxies which scatter further from the power law relation evident in the compact population. This could be tantalising observational evidence for the galaxies that populate the diffuse population.

\begin{figure*}
  \includegraphics[width=\linewidth]{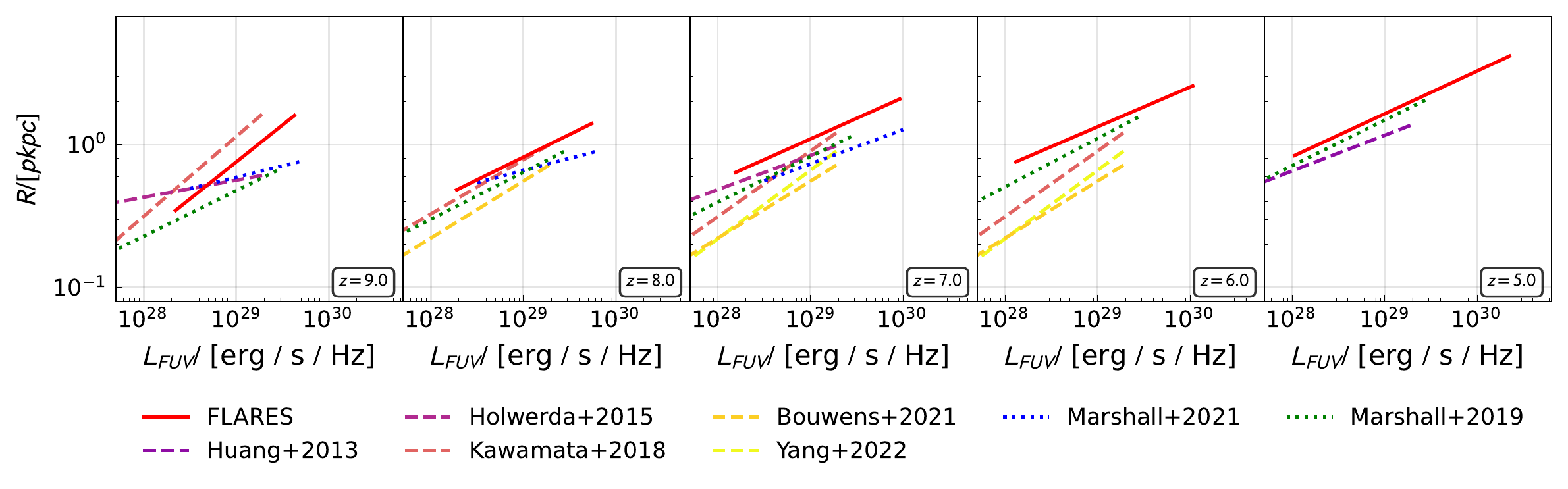}
  \caption{Fits to the UV size-luminosity relation including the effects of dust measured using the pixel method. To perform the fits we use the entire complete galaxy sample. We include comparisons to observations without lensed galaxies \citep{Huang_2013, Holwerda_2015}, observations including lensed sources \citep{Kawamata_2018, Bouwens2021, Yang2022} and simulations \citep{Marshall2019, Marshall21}. We denote \flares\ by a red solid line, observations by dashed lines and other simulations by dotted lines.}
  \label{fig:fits_obscomp}
\end{figure*}

\begin{table*}
\begin{center}
\begin{tabular}{ c | c | c | c }
 \hline
 Redshift ($z$) 
 & $R_0/ [\mathrm{pkpc}]$ & $\beta$ \\ \hline
 9 
 & 0.793$\pm$0.019 & 0.519$\pm$0.026 \\
 8 
 & 0.842$\pm$0.012 & 0.319$\pm$0.013 \\  
 7 
 & 1.126$\pm$0.011 & 0.290$\pm$0.008 \\
 6 
 & 1.370$\pm$0.007 & 0.279$\pm$0.004 \\
 5 
 & 1.692$\pm$0.006 & 0.300$\pm$0.003 \\
 \hline
\end{tabular}
\caption{The fitting results for \eq{size_lumin_fit} for each redshift bin in \fig{fits_obscomp} for the attenuated size-luminosity relations, measured using the pixel method (\sec{pixel}). $R_0$ is a normalisation factor, and $\beta$ is the slope of the size-luminosity relation.}
\label{tab:rfit}
\end{center}
\end{table*}

To quantify the agreement between the observational scatter and the \flares\ sample we use \texttt{curve\_fit} (non-linear least squares fitting), from \texttt{scipy} \citep{2020SciPy-NMeth}, to produce fits of the form of \eq{size_lumin_fit}. The results of this fitting are shown in \tab{rfit}.

\fig{fits_obscomp} shows a comparison of these fits (solid red lines) to fits from observed samples: \cite{Huang_2013} at $z=5$, \cite{Holwerda_2015} at $z=7$ and $z=9$, \cite{Kawamata_2018} at $z=6-9$, \cite{Bouwens2021} at $z=6-8$, and \cite{Yang2022} at $z=6-7$, the latter 3 of these including lensed sources. We also compare to two simulations: the \textsc{Meraxes} semi-analytic model \citep{Liu_meraxes, Marshall2019} at $z=5-9$, and the \bluetides\ simulation \citep{Marshall21} at $z=7-9$. We denote observations with dotted lines and simulations (other than \flares) by dotted lines. Each fit is plotted using their published fitting parameters.

At $z>7$ the \flares\ fits exhibit a good agreement in slope with the observational studies including lensed samples. These fits are significantly steeper than the observational samples that do not have a contribution of lensed galaxies, as demonstrated in \cite{Bouwens2021}. At $z\leq7$ the \flares\ fits begin to flatten relative to the studies including lensed sources as galaxies in the dim and diffuse size-luminosity regime become more numerous. 

Compared to \bluetides, we find \flares\ has a steeper size-luminosity relation at $z=8-9$ and a stronger redshift evolution in the normalisation over the redshift range $7\leq z \leq9$. With respect to \textsc{Meraxes} we find a good agreement in slopes at $z<9$ with a consistently higher normalisation at all redshifts.

Each work predicts a different normalisation of the size-luminosity relation. This is particularly evident at $z<8$ where \flares\ has consistently higher normalisation than all other studies. One explanation for this difference is the resolution and measurement methods in each study. The pixel method used in this work is sensitive to the resolution of the image (for which we adopt the softening length of the simulation), observational studies on the other hand use images with a higher resolution than the softening length of \flares\ and use an array of measurement techniques that are less sensitive to the pixel resolution. \bluetides\ uses the pixel method but adopts a higher pixel resolution below the softening length of the simulation and \textsc{Meraxes} derive their sizes (scale radius of the disc) from the SAM galaxy properties. In addition to methodological differences, there is likely a significant contribution to the normalisation by the diffuse galaxies, which at fixed luminosity extend to larger sizes in the \flares\ sample. 

The slopes reported in \tab{rfit} for the attenuated size-luminosity relation are in broad agreement with the results of \cite{Grazian_2012}, \cite{Huang_2013}, \cite{Holwerda_2015}, \cite{Shibuya2015}, \cite{Kawamata_2018}, \cite{Bouwens2021} and \cite{Yang2022} in various different redshift regimes. At $z>7$ the \flares\ results exhibit the steeper slopes present in \cite{Kawamata_2018}, \cite{Bouwens2021}, and \cite{Yang2022} before flattening into closer agreement with \cite{Grazian_2012}, \cite{Huang_2013}, \cite{Holwerda_2015} and \cite{Shibuya2015} at $z\leq7$. Again, this is due to the aforementioned compact low luminosity galaxies present in the lensed samples, which are absent from the other studies, and the diffuse low luminosity galaxies in the \flares\ sample which become more numerous with decreasing redshift. 

Many of the compact galaxies that strongly affect the slope of the size-luminosity relation in lensing studies fall below the resolution limit of \flares\ (indicated by the dashed line in \fig{HLR_with_dust_obscomp}) and \bluetides. Higher resolution simulations are necessary to ascertain if these galaxies are present in the simulated sample and produce the same steepening behaviour. All observational samples also lack the most diffuse galaxies in the simulated samples due to their low surface densities. These would act to flatten the size-luminosity relation if present. Future works will aim to address both these issues with higher resolution simulations and fully synthetic observations including survey limits, instrument noise, point spread functions and observational methods of structure detection; the former addressing the missing dim and compact galaxies in the simulated sample and the latter addressing the diffuse galaxies which are likely undetected in the observational sample.

\subsection{The size-luminosity relation as a function of wavelength}

\begin{figure*}
    \centering
	\includegraphics[width=\linewidth]{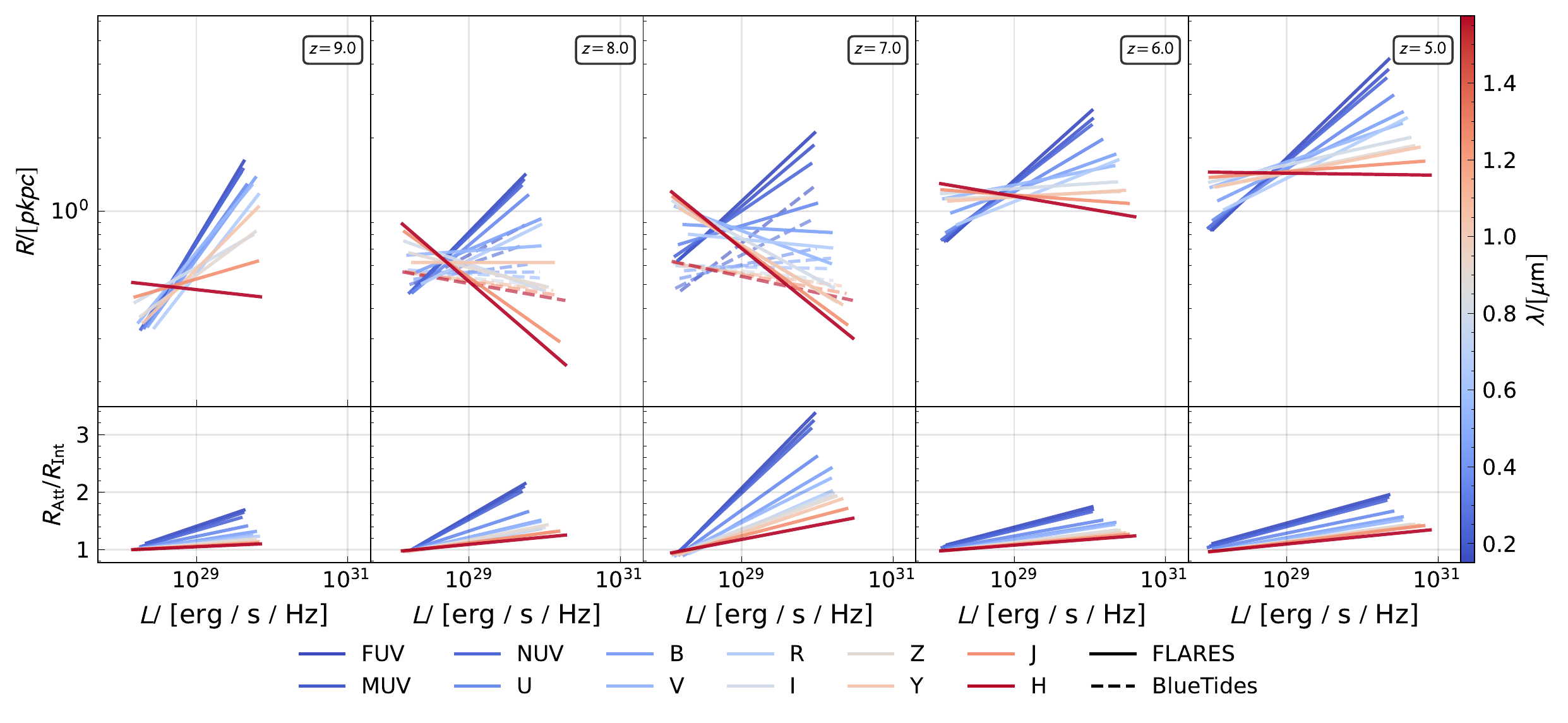}
    \caption{The upper row of panels show fits to the size-luminosity relation for all rest frame bands in \fig{example_sed}. Solid lines represent the \flares\ fits while dashed lines show the \bluetides\ \citep{Marshall21} fits for the same selection of bands. The lower row of panels show straight line fits to the ratio between the intrinsic and attenuated sizes for each band. The colour of the line denotes the band, with the bluest bands in blue and reddest bands in red. The colorbar shows the central wavelength of each rest frame band in microns.}
    \label{fig:colors}
\end{figure*}

In \fig{colors} we present the size luminosity relation across a range of rest frame filters (shown in \fig{example_sed}), and compare to the corresponding fits from \cite{Marshall21} at $z=[8, 7]$. We present the fitting parameters in \app{bandtable}.

As the probed wavelength regime reddens, the slope of the size-luminosity relation decreases, becoming increasingly negative for the reddest filters. These red filters probe the underlying stellar distribution with the least attenuation. The increasing representation of the underlying intrinsic distribution is clearly shown in the bottom row of panels as the slope of the ratio between attenuated and intrinsic size flattens with increasing wavelength. The slope of the size-luminosity relation for the reddest filters increases with decreasing redshift, implying that the intrinsic stellar population is becoming more diffuse as galaxies evolve. 

This variation with wavelength is also predicted by \bluetides\ \citep{Marshall21} at $z=[7,8]$, although they predict a shallower size-luminosity relation for the reddest filters relative to those produced in this work. It is also consistent with observations at low redshift \citep[e.g.][]{Barbera10, Kelvin12, Vulcani14, Kennedy15, Tacchella_2015}. 

Nonetheless, these results present a tantalising prediction which will allow Webb to ascertain the validity of the negative intrinsic size-luminosity relation. Webb's reddest broad-band NIRCam filter (F444W) will probe as blue as the B band at $z=9$ and I at $z=5$ (as shown in \fig{example_sed}) allowing for high resolution measurements of galaxy sizes in this regime.

\subsection{Redshift Evolution}

\begin{figure}
	\includegraphics[width=\columnwidth]{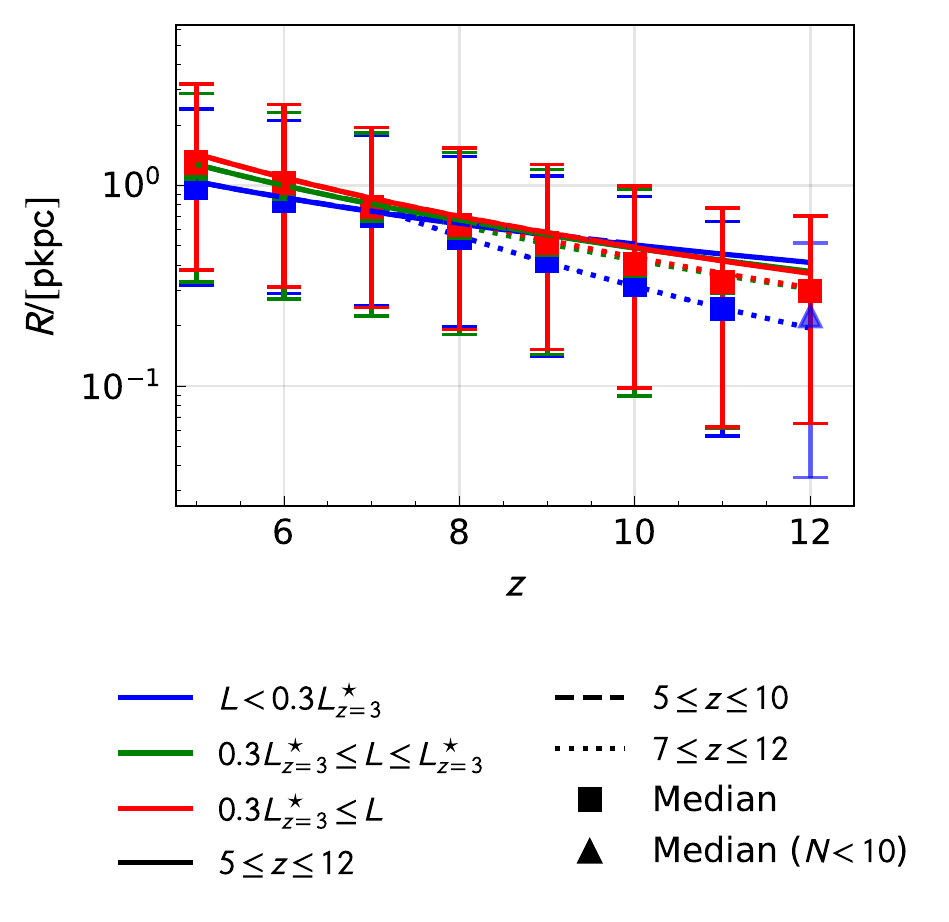}
    \caption{The redshift evolution of galaxy size in the \flares\ sample split into 3 luminosity samples used in the literature: a low luminosity sample, $L < 0.3 \, L_{z=3}^{\star}$ (blue), an intermediate luminosity sample, $0.3 \, L_{z=3}^{\star} < L < L_{z=3}^{\star}$ (green), and a bright galaxy sample, $0.3 \, L_{z=3}^{\star}<L$ (red), where $L_{z=3}^{\star}\approx10^{29}$ erg s$^{-1}$ Hz$^{-1}$. We present 3 different fits the different redshift regimes: a solid line fit to the entire redshift range ($5 \leq z \leq 12$), a dashed line fit to a low-$z$ sample ($5 \leq z \leq 10$) and a dotted line fit to a high redshift sample ($7 \leq z \leq 12$). The low-$z$ fits and the full redshift range fits almost entirely overlap.
    The points show the median in each redshift bin with errorbars denoting the 16th and 84th percentile, a square point denotes more than 10 galaxies in the bin while a triangle denotes less than 10 galaxies present in the sample at that redshift.}
    \label{fig:evo}
\end{figure}

\begin{figure}
    \includegraphics[width=\columnwidth]{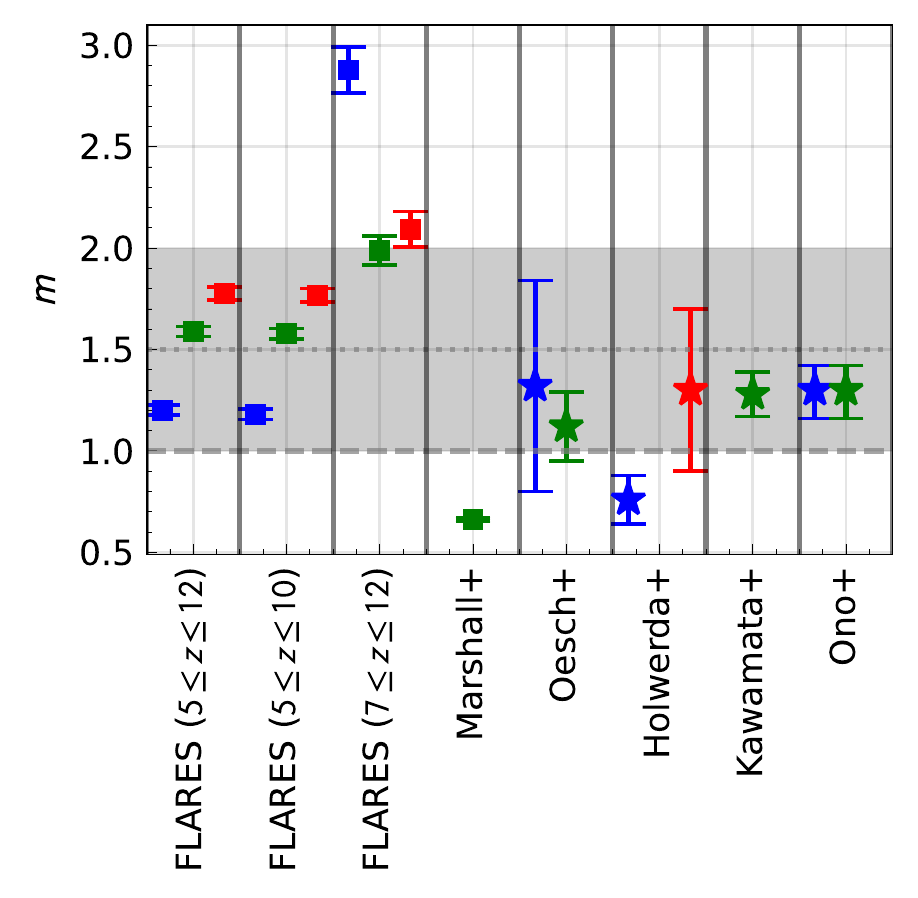}
    \caption{A comparison to the slopes of the redshift size evolution derived from observations \citep{Oesch_2010, Ono_2013, Holwerda_2015, Kawamata_2018}, and the \bluetides\ simulation \citep{Marshall21}. Observations are denoted by stars and simulations are denoted by squares. The shaded range shows the range of slopes found in the FIRE-2 simulations \citep{Ma_18_size} for various fixed mass and luminosity galaxy samples. The dashed line corresponds to $m=1$, the theoretical scaling for systems of fixed mass \citep[e.g.][]{Bouwens_2004}, and the dotted line corresponds to $m=1.5$, the theoretical scaling for systems with fixed circular velocity \citep[e.g.][]{Ferguson_2004, Hathi_2008}. As with \fig{evo}, blue points represent a low luminosity sample ($L < 0.3 \, L_{z=3}^{\star}$), green points represent an intermediate luminosity sample ($0.3 \, L_{z=3}^{\star} < L < L_{z=3}^{\star}$), and red points represent a bright galaxy sample ($0.3 \, L_{z=3}^{\star}<L$).}
    \label{fig:evoslopes}
\end{figure}

In the literature there has been a wide range of presented methods for measuring the redshift evolution of galaxy sizes, with various approaches and galaxy sample definitions used for the computation. To produce a comprehensive comparison with \flares\ we employ non-linear least squares fitting (again using \texttt{scipy.curve\_fit}) to produce fits to \eq{evo} from various sample definitions pulled from the complete galaxy sample, all weighted with the \flares\ weighting scheme. The results of this fitting are presented in \tab{evo}.

In \fig{evo} we present these fits for a number of different sample definitions found in the literature. \fig{evoslopes} shows a comparison of the slope ($m$) from various studies, left to right: \flares, \cite{Marshall21}, \cite{Oesch_2010}, \cite{Holwerda_2015}, \cite{Kawamata_2018}, and \cite{Ono_2013}, with a shaded region representing the range of slopes from \cite{Ma_18_size}. We present the fitting parameters for these fits in \tab{evo}.

For the low luminosity sample we see a good agreement in slope between \flares\ and \cite{Oesch_2010} and \cite{Ono_2013}. For the other \flares\ samples we find comparatively high slopes compared to the other works. However, these values are in agreement with \cite{Ma_18_size} who predict values in the range $m=1-2$ depending on the fixed mass or luminosity regime (shown by the shaded region). All but the low luminosity sample's slopes are larger than the evolution of systems at fixed circular velocity, implying an increasing feedback contribution to the evolution with decreasing redshift. Conversely, the low luminosity sample's evolution is closer to that of a system at fixed mass with the same additional feedback contribution. As feedback becomes more efficient with decreasing redshift the star forming gas will be given more thermal energy and thus change the dynamics of the star forming gas, increasing the radii at which stars can form and thus the half light radii.


Limiting the included redshifts in the \flares\ sample can not only be used to compare to the more limited samples of \bluetides, with no galaxies at $z<7$, and observations, where $z\geq10$ galaxies are exceedingly rare, but can also probe the evolution of size during particular epochs. To do this we limited the sample to a high-$z$ sample limited to $z\geq7$ and a low-$z$ sample with $z\leq10$, the results of which are also included in \tab{evo}. Limiting to $z\geq7$ resulted in a large increase in the slope of the redshift evolution alongside unrealistically high normalisations, predicting $z=0$ sizes of the order $\sim300$ pkpc for the low luminosity sample and over double the $z=0$ size in the limited and high luminosity samples in the other redshift selections. Conversely, limiting to $z\leq10$ instead results in fitting results consistent with those produced by the full redshift range. This casts doubt on the sparse $z>10$ measurements in observations causing the differences in slope between the \flares\ measurement and observational measurements. More interestingly the differences in fits between redshift regimes implies a significantly faster evolution of galaxy size at the earliest times, even for the most dim and diffuse galaxies in the low luminosity sample. It is clear from \fig{evo} that a piecewise fit produces a considerably better fit to the data than fitting across the entire redshift range. 

Tensions between \flares\ and the observations are far less stark than those between \flares\ and \bluetides\ samples but are nonetheless evident for the capped and high luminosity samples, we do however see a good agreement in the low luminosity sample. The tensions here could be explained by how sparse observations are at the highest redshifts due to the small area covered at the required depth; given that the low luminosity sample in \flares\ is also sparse at the highest redshifts, the agreement between observations and \flares\ here could be due to this luminosity regime being where the simulation and observations have the largest overlap in sampling strength. Additional observations from upcoming observatories populating the highest redshifts will increase the area and depth sampled in at this epoch and could rectify this tension. It should also be noted however that subgrid models require intensive investigation at this epoch, with comparison to robust observations to ascertain the validity of their behaviour. Future work will be able to converge the results of both simulations and observations to a consistent story of galaxy size evolution.

\begin{table*}
\begin{center}
\begin{tabular}{ c | c | c | c | c | c | c | c }
 \hline
   & \multicolumn{2}{|c|}{$5 \leq z \leq 12$} & \multicolumn{2}{|c|}{$5 \leq z \leq 10$} & \multicolumn{2}{|c|}{$7 \leq z \leq 12$} \\ \hline
 Sample & $R_{0,z=0}/ [\mathrm{pkpc}]$ & $m$ & $R_{0,z=0}/ [\mathrm{pkpc}]$ & $m$ &  $R_{0,z=0}/ [\mathrm{pkpc}]$ & $m$ \\ \hline
 $L<0.3L_{z=3}^{\star}$ 
 & 8.99$\pm$0.42  & 1.20$\pm$0.03
 & 8.65$\pm$0.41  & 1.18$\pm$0.03
 & 311.78$\pm$73.80 & 2.88$\pm$0.11\\  
 $0.3L_{z=3}^{\star}<L<L_{z=3}^{\star}$ 
 & 21.98$\pm$1.04  & 1.59$\pm$0.03
 & 21.53$\pm$1.04   & 1.58$\pm$0.03
 & 49.91$\pm$7.62  & 1.99$\pm$0.07 \\ 
 $0.3L_{z=3}^{\star}<L$ 
 & 34.61$\pm$2.08  & 1.78$\pm$0.03
 & 34.11$\pm$2.09  & 1.77$\pm$0.03
 & 66.22$\pm$12.43  & 2.09$\pm$0.09\\ 
 \hline
\end{tabular}
\caption{The fitting parameters for \eq{evo} in \fig{evo} split into 3 redshift samples. From left to right: the full \flares\ sample, a sample excluding the highest redshifts where robust observations are sparse, and a sample excluding the lowest redshift snapshots for comparison to \bluetides. $R_{0,z=0}$ is a normalisation factor corresponding to a galaxy's size at $z=0$ and $m$ is the slope of the redshift evolution.}
\label{tab:evo}
\end{center}
\end{table*}

\section{Conclusions}
\label{sec:conclusion}

In this paper we have presented an analysis of galaxy sizes at $z\geq5$ in the \flares\ simulations across a wide array of environments. To do this we produced synthetic galaxy images using photometry in rest frame UV and visual bands derived using the line of sight attenuation method presented in \cite{Vijayan2020}. We presented an efficient method of image computation by utilising a KD-Tree of pixel coordinates and smoothing stellar particles over their SPH kernels. We employed this imaging method to produce synthetic galaxy images, from which the size of galaxies were measured using a non-parametric pixel based method to account for the clumpy nature of galaxies at high redshift. 

Using these measurements we probed both the intrinsic and observed size-luminosity relation in the rest frame far-UV (1500 \AA), finding:

\begin{itemize}
    \item The intrinsic size-luminosity relation is bi-modal, with one intrinsically compact and bright population and one intrinsically diffuse and dim population.
    \item These 2 populations result in a negative slope to the rest-frame far-UV intrinsic size-luminosity distribution.
    \item Including the effects of dust attenuation results in the perceived size of galaxies to increase, with the most intrinsically compact galaxies increase in size by as much as $\times50$. 
    \item The increase in size due to dust attenuation inverts the slope of the size-luminosity relation, resulting in a fair agreement between observations and in this work. However, the \flares\ sample lacks low luminosity compact galaxies which have been shown to steepen the size-luminosity relation in lensing studies. Conversely, the observational samples lack the diffuse and dim galaxies that are present in this work, these act to flatten the size-luminosity relation. The affects of these missing galaxies highlights the need for high resolution simulations in the future and observationally motivated measurement methods.
    \item Dust distributions in these compact galaxies are highly concentrated with half metal radii of $<1$ pkpc, heavily attenuating the intrinsically bright cores and increasing the observed half light radius. This may be observable as strong dust gradients. 
\end{itemize}

We performed size measurements for a range of rest frame UV and visual bands, finding an anti-correlation between the slope of the size-luminosity relation and wavelength. This anti-correlation becomes weaker with decreasing redshift as the intrinsic stellar distribution increases in size. This represents a falsifiable prediction which Webb will be able to probe at high resolution with NIRCam.

We then investigated the evolution of size with redshift in the far-UV, finding slopes for multiple sample definitions in the range $m=1.21-1.87$. These values are consistent with theoretical predictions modified by additional contributions to the evolution by feedback mechanisms. At low luminosity the evolution is consistent with an evolution at fixed mass ($m=1$) with additional evolution due to feedback, while high luminosity galaxies are consistent with a fixed circular velocity evolution ($m=1.5$), again with an additional contribution from feedback. 
With the exception of the low luminosity sample giving a good agreement, these results are in tension with observations. They do however broadly agree with the range found in the FIRE-2 simulations. The limited observational galaxy sample at extremely high redshifts could contribute to this tension. 
Limiting the galaxy sample to both a low ($5 \leq z \leq 10$) and high ($7 \leq z \leq 12$) redshift sample yielded little change in the results for the low redshift sample but resulted in significantly higher slopes for the high redshift sample. This implies a non-constant size evolution with faster evolution in the highest redshift bins. Further observations from future high redshifts surveys are needed to probe the differences highlighted here in addition to future simulations adding to the theory. 


With the launch of Webb we will soon be able to probe these high redshift regimes with far greater fidelity and further strengthen our understanding of the earliest epochs of galaxy evolution. Webb will allow us to probe higher redshifts at high resolution with NIRCam. Not only will this further populate galaxy samples at $z>8$, it will also increase the completeness of the high redshift observational surveys at low luminosity.

Future work will include the next generation of \flares\ simulating a wider range of environments, probing more regions, and simulating a significant volume at high mass resolution. Including higher resolution simulations will enable comparison to the dim and compact galaxies found in lensing studies, while increasing the effective volume with more resimulated regions will allow \flares\ to reach a volume comparable to the largest upcoming observational surveys from \euclid.   

In addition to the next generation of \flares, the underlying physical processes governing the size evolution in the subgrid model will be probed. This will include stellar and AGN feedback, star formation conditions and chemical enrichment. The effects of simulation and observational structure detection methods will be investigated to quantify the effect of survey depth and the segmentation of substructures. In particular this will aim to probe the effects of structure detection methods on the diffuse galaxy population and the effect this has on the size-luminosity relation.

\section*{Acknowledgements}

We thank the \eagle\ team for their efforts in developing the \eagle\ simulation code.

We acknowledge the indispensable contribution from the publicly available programming language \texttt{python} \citep{Python}, including the \texttt{numpy} \citep{numpy}, \texttt{astropy} \citep{Astropy}, \texttt{matplotlib} \citep{Matplotlib}, \texttt{scipy} \citep{2020SciPy-NMeth}, and \texttt{h5py} \citep{h5py} packages. 

This work used the DiRAC@Durham facility managed by the Institute for Computational Cosmology on behalf of the STFC DiRAC HPC Facility (www.dirac.ac.uk). The equipment was funded by BEIS capital funding via STFC capital grants ST/K00042X/1, ST/P002293/1, ST/R002371/1 and ST/S002502/1, Durham University and STFC operations grant ST/R000832/1. DiRAC is part of the National e-Infrastructure.

CCL acknowledges support from the Royal Society under grant RGF/EA/181016. DI acknowledges support by the European Research Council via ERC Consolidator Grant KETJU (no. 818930). The Cosmic Dawn Center (DAWN) is funded by the Danish National Research Foundation under grant No. 140. MAM acknowledges the support of a National Research Council of Canada Plaskett Fellowship, and the Australian Research Council Centre of Excellence for All Sky Astrophysics in 3 Dimensions (ASTRO 3D), through project number CE170100013.

\section*{Data Availability}
The integrated galaxy properties used to generate the plots in this article is available at \href{https://flaresimulations.github.io/data.html}{flaresimulations.github.io/data}. More detailed data, including particle data, can be provided upon request. All the code used to produce the analysis in this article is public and available at \href{https://github.com/WillJRoper/flares-sizes-obs}{github.com/WillJRoper/flares-sizes-obs}.




\bibliographystyle{mnras}
\bibliography{flaresIV} 




\appendix

\section{The effects of smoothing}
\label{sec:smooth}

Here we present comparisons between smoothing methods used in image creation first comparing Gaussian and spline kernel smoothing and then the differences between smoothing and ignoring smoothing.

\subsection{Comparing kernel averaging to Gaussian smoothing}

\begin{figure*}
	\includegraphics[width=\linewidth]{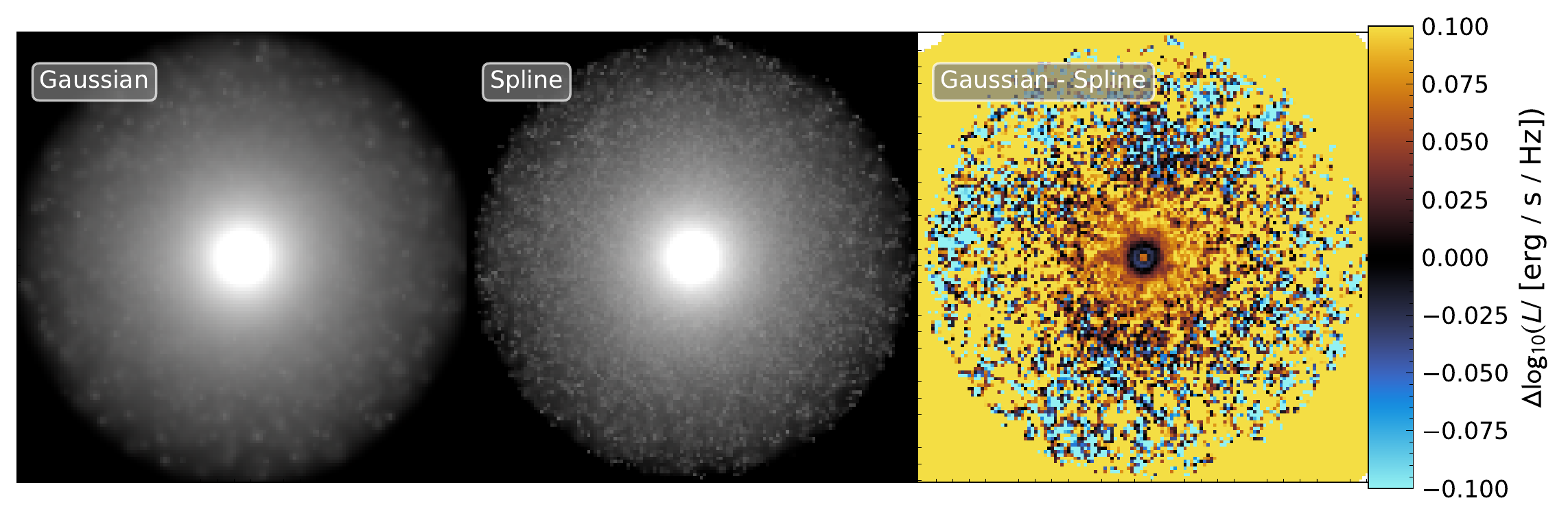}
    \caption{A comparison of logscaled stacked images produced using the Gaussian smoothing method (left), spline kernel method (middle) and a residual image showing the difference between the log of the two methods images. The images themselves are stacks in the far UV of all galaxies in the \flares\ sample (irrespective of completeness).}
    \label{fig:imgtyperesi}
\end{figure*}

\fig{imgtyperesi} shows a comparison between the Gaussian and spline smoothing methods. Qualitatively it can be seen the Gaussian method results in a smoother light distribution due to the indefinite boundaries of the Gaussian smoothing kernel, this spreads light beyond the `extent' given by the SPH kernel. The spline method produces a more granular image with clearer small structures at the outskirts of the FOV. The residual image shows that the Gaussian method's spreading of light leads to differences at large radii where the Gaussian image is brighter due to the spreading of light. However, this does not mean the Gaussian image is consistently more luminous at large radii, compact structures at large radii in the spline image have more concentrated emission causing these regions to out shine the Gaussian image. This effects is also noticeable in the centre of the image where there is a ring of spline dominated pixels due to this concentration of light. These effects are however minimal with each image differing at most by 0.1 dex. 

\begin{figure}
	\includegraphics[width=\columnwidth]{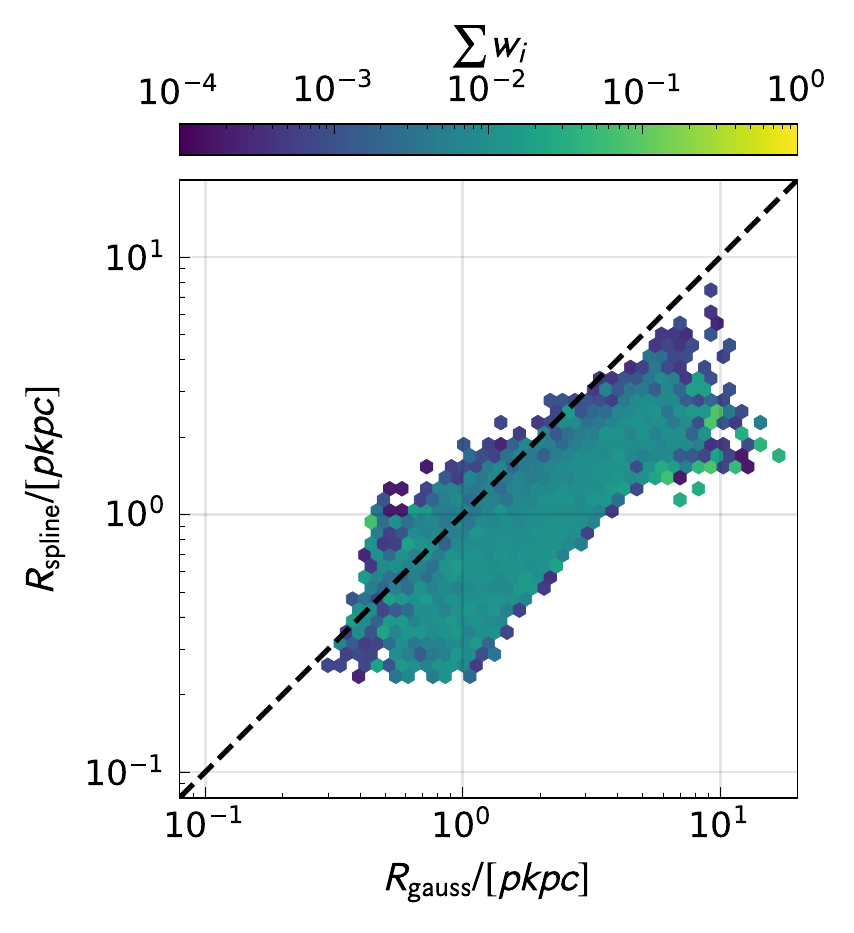}
    \caption{A comparison between the sizes of galaxies measured using the pixel method from the spline (y-axis) and Gaussian (x-axis) smoothing methods. The dashed line represents a 1:1 relation. In this plot we do not differentiate between the compact and diffuse galaxy populations and only present the full complete sample.}
    \label{fig:imgtypehlr}
\end{figure}

We further show the effects of smoothing method in \fig{imgtypehlr} where we compare the measured sizes of galaxies in each method. In the vast majority of cases the Gaussian smoothing results in a larger perceived size due to the increased spread of a single stellar particle's luminosity. The instances where the spline method yields larger sizes are dominated by smaller galaxies where the dilution of the Gaussian method causes structures to occupy more pixels relative to the more concentrated spline method and thus a larger area is used in the pixel driven size calculation. It should be noted here that the spline method produces a better agreement with observations with the Gaussian method producing size-luminosity relations which overestimate galaxy sizes relative to observations.

\subsection{Smoothing vs no smoothing}

\begin{figure}
	\includegraphics[width=\columnwidth]{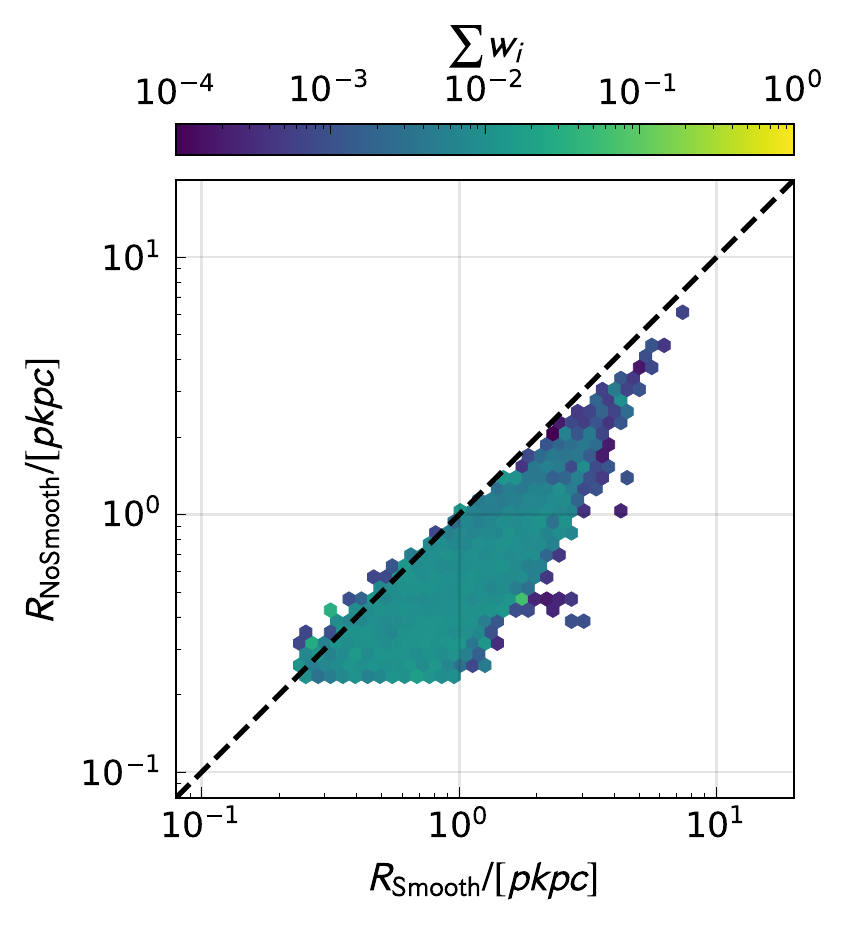}
    \caption{A comparison between sizes of galaxies measured using the pixel method from dust attenuated images with and without smoothing of the stellar particles. The dashed line represents a 1:1 relation. In this plot we do not differentiate between the compact and diffuse galaxy populations and only present the full complete sample.}
    \label{fig:nosmooth}
\end{figure}

In \fig{nosmooth} we compare the spline smoothing method to galaxy sizes measured from images where no smoothing has been performed on the stellar particles. In some cases there is minimal difference between the smoothed and unsmoothed measurements, particularly for compact galaxies where the stellar kernels themselves are very small resulting in minimal smoothing. In the vast majority of cases the smoothing increases the measured size, with the most diffuse incomplete galaxies (transparent distribution) extending to much larger sizes when smoothed.

\section{Size-luminosity relation wavelength variation}

In this appendix we present the fitting parameters for the wavelength evolution of the size-luminosity relation shown in \fig{colors}.

\label{sec:bandtable}

\begin{table*}
\begin{center}
\begin{tabular}{ c | c | c | c | c | c | c | c | c | c }
 \hline
 Redshift ($z$) & \multicolumn{2}{|c|}{9} & \multicolumn{2}{|c|}{8} & \multicolumn{2}{|c|}{7} \\ \hline
 Band & $R_0$ & $\beta$& $R_0$ & $\beta$ & $R_0$ & $\beta$ \\ \hline
 FUV 
 & 0.793 +/- 0.019 & 0.519 +/- 0.026
 & 0.842 +/- 0.012 & 0.319 +/- 0.013
 & 1.126 +/- 0.011 & 0.290 +/- 0.008 \\
 MUV 
 & 0.773 +/- 0.020 & 0.493 +/- 0.026
 & 0.821 +/- 0.012 & 0.313 +/- 0.013
 & 1.070 +/- 0.011 & 0.263 +/- 0.008 \\
 NUV 
 & 0.777 +/- 0.021 & 0.485 +/- 0.026
 & 0.813 +/- 0.013 & 0.296 +/- 0.014 
 & 1.020 +/- 0.013 & 0.211 +/- 0.009 \\
 U 
 & 0.687 +/- 0.017 & 0.434 +/- 0.026 
 & 0.743 +/- 0.011 & 0.262 +/- 0.014 
 & 0.878 +/- 0.011 & 0.092 +/- 0.010 \\
 B 
 & 0.660 +/- 0.014 & 0.428 +/- 0.025 
 & 0.704 +/- 0.010 & 0.133 +/- 0.014 
 & 0.854 +/- 0.010 & -0.017 +/- 0.010 \\
 V 
 & 0.702 +/- 0.018 & 0.375 +/- 0.024 
 & 0.689 +/- 0.013 & 0.022 +/- 0.014 
 & 0.823 +/- 0.011 & -0.114 +/- 0.009 \\
 R
 & 0.573 +/- 0.011 & 0.397 +/- 0.027 
 & 0.638 +/- 0.008 & 0.154 +/- 0.014
 & 0.765 +/- 0.009 & -0.030 +/- 0.011 \\
 I
 & 0.598 +/- 0.019 & 0.178 +/- 0.026 
 & 0.601 +/- 0.013 & -0.110 +/- 0.015 
 & 0.763 +/- 0.011 & -0.167 +/- 0.009 \\
 Z
 & 0.558 +/- 0.015 & 0.229 +/- 0.027 
 & 0.583 +/- 0.011 & -0.077 +/- 0.015 
 & 0.715 +/- 0.010 & -0.186 +/- 0.010 \\
 Y
 & 0.595 +/- 0.014 & 0.312 +/- 0.026 
 & 0.616 +/- 0.010 & -0.000 +/- 0.014 
 & 0.715 +/- 0.010 & -0.183 +/- 0.010 \\
 J
 & 0.532 +/- 0.017 & 0.090 +/- 0.027 
 & 0.525 +/- 0.011 & -0.220 +/- 0.015 
 & 0.698 +/- 0.010 & -0.228 +/- 0.009 \\
 H
 & 0.476 +/- 0.017 & -0.035 +/- 0.027 
 & 0.503 +/- 0.011 & -0.268 +/- 0.014
 & 0.688 +/- 0.010 & -0.250 +/- 0.008 \\
 \hline
\end{tabular}
\caption{The fitting results for \eq{size_lumin_fit} for $z=7-9$ and all rest frame bands in \fig{colors}. $R_0$ is a normalisation factor, $\beta$ is the slope of the size-luminosity relation and $N$ is the number of galaxies used in each fit.}
\label{tab:colorrfit1}
\end{center}
\end{table*}

\begin{table*}
\begin{center}
\begin{tabular}{ c | c | c | c | c | c | c}
 \hline
 Redshift ($z$) & \multicolumn{2}{|c|}{6} & \multicolumn{2}{|c|}{5} \\ \hline
 Band & $R_0$ & $\beta$ & $R_0$ & $\beta$ \\ \hline
 FUV 
 & 1.370 +/- 0.007 & 0.279 +/- 0.004 
 & 1.692 +/- 0.006 & 0.300 +/- 0.003 \\
 MUV 
 & 1.326 +/- 0.007 & 0.256 +/- 0.004
 & 1.639 +/- 0.006 & 0.280 +/- 0.003 \\
 NUV 
 & 1.315 +/- 0.008 & 0.238 +/- 0.004
 & 1.627 +/- 0.006 & 0.261 +/- 0.003 \\
 U 
 & 1.218 +/- 0.007 & 0.184 +/- 0.004
 & 1.514 +/- 0.006 & 0.215 +/- 0.003 \\
 B 
 & 1.227 +/- 0.007 & 0.111 +/- 0.004 
 & 1.526 +/- 0.005 & 0.149 +/- 0.003 \\
 V 
 & 1.285 +/- 0.008 & 0.060 +/- 0.004 
 & 1.604 +/- 0.006 & 0.104 +/- 0.002 \\
 R
 & 1.106 +/- 0.005 & 0.124 +/- 0.005 
 & 1.383 +/- 0.004 & 0.156 +/- 0.003 \\
 I
 & 1.238 +/- 0.008 & 0.021 +/- 0.004 
 & 1.554 +/- 0.006 & 0.069 +/- 0.002 \\
 Z
 & 1.155 +/- 0.007 & 0.013 +/- 0.004 
 & 1.455 +/- 0.005 & 0.064 +/- 0.002 \\
 Y
 & 1.143 +/- 0.007 & 0.019 +/- 0.004 
 & 1.439 +/- 0.005 & 0.061 +/- 0.002 \\
 J
 & 1.161 +/- 0.007 & -0.023 +/- 0.004 
 & 1.455 +/- 0.005 & 0.024 +/- 0.002 \\
 H
 & 1.146 +/- 0.007 & -0.053 +/- 0.004
 & 1.430 +/- 0.005 & -0.004 +/- 0.002 \\
 \hline
\end{tabular}
\caption{The fitting results for \eq{size_lumin_fit} for $z=6-5$ and all rest frame bands in \fig{colors}. $R_0$ is a normalisation factor, $\beta$ is the slope of the size-luminosity relation and $N$ is the number of galaxies used in each fit.}
\label{tab:colorrfit2}
\end{center}
\end{table*}


\bsp	
\label{lastpage}
\end{document}